\documentclass[prd,twocolumn,showpacs,showkeys,preprintnumbers,floatfix,nofootinbib,superscriptaddress]{revtex4-1}
\usepackage{amsfonts} 
\usepackage{amssymb} 
\usepackage{amsmath} 
\usepackage{graphicx} 
\usepackage{subfigure} 
\usepackage{array} 
\usepackage{dcolumn} 
\usepackage{bm} 
\usepackage{latexsym} 
\usepackage{longtable} 
\usepackage{hyperref} 
\usepackage{bbold}
\usepackage{color}
\usepackage{multirow}
\usepackage{rotating}

\graphicspath{{./figs/}}

\newcommand{\qslash}[1]{\text{$\not \! #1$}}

\newcommand{\MeV}{\mathop{\rm MeV}\nolimits}
\newcommand{\GeV}{\mathop{\rm GeV}\nolimits}
\newcommand{\fm}{\mathop{\rm fm}\nolimits}

\DeclareMathOperator{\Tr}{Tr}

\newcommand{\slantlabel}[1]{\hspace{0.15cm}\begin{rotate}{30}{#1}\end{rotate}\hspace{-0.15cm}}
\definecolor{green}{rgb}{0.1, 0.8, 0.1}

\def\good{\raisebox{1.2pt}{\makebox[10pt]{\color{green}\small$\bigstar$}}}
\def\soso{\makebox[10pt]{\color{green}\Large$\circ$}}
\def\bad{\raisebox{1pt}{\makebox[10pt]{\color{red}\footnotesize$\blacksquare$}}}

\begin{document}


\title{Isovector and Isoscalar Tensor Charges of the Nucleon from Lattice QCD}
\author{Tanmoy Bhattacharya}
\email{tanmoy@lanl.gov}
\affiliation{Los Alamos National Laboratory, Theoretical Division T-2, Los Alamos, NM 87545}

\author{Vincenzo Cirigliano}
\email{cirigliano@lanl.gov}
\affiliation{Los Alamos National Laboratory, Theoretical Division T-2, Los Alamos, NM 87545}

\author{Saul D. Cohen}
\email{saul.cohen@gmail.com}
\affiliation{Institute for Nuclear Theory, University of Washington, Seattle, WA 98195}

\author{Rajan Gupta}
\email{rajan@lanl.gov}
\affiliation{Los Alamos National Laboratory, Theoretical Division T-2, Los Alamos, NM 87545}

\author{Anosh Joseph}
\email{anosh.joseph@desy.de}
\affiliation{John von Neumann Institute for Computing, DESY, 15738 Zeuthen, Germany}

\author{Huey-Wen Lin}
\email{hueywenlin@lbl.gov}
\affiliation{Physics Department, University of California, Berkeley, CA 94720}

\author{Boram Yoon}
\email{boram@lanl.gov}
\affiliation{Los Alamos National Laboratory, Theoretical Division T-2, Los Alamos, NM 87545}

\collaboration{Precision Neutron Decay Matrix Elements (PNDME) Collaboration}
\preprint{LA-UR-15-23801}
%
\date{\today}
\begin{abstract}
We present results for the isovector and flavor diagonal tensor charges
$g^{u-d}_T$, $g^{u}_T$, $g^{d}_T$, and
$g^{s}_T$ needed to probe novel tensor interactions at the TeV scale
in neutron and nuclear $\beta$-decays and the contribution of the
quark electric dipole moment (EDM) to the neutron EDM.
The lattice QCD calculations were done using nine ensembles of gauge
configurations generated by the MILC collaboration using the HISQ
action with 2+1+1 dynamical flavors. These ensembles span three
lattice spacings $a \approx 0.06, 0.09$ and $0.12 \fm$ and three quark
masses corresponding to the pion masses $M_\pi \approx 130, 220$ and
$310 \MeV$. Using estimates from these ensembles, we quantify all 
systematic uncertainties and perform a simultaneous extrapolation in
the lattice spacing, volume and light quark masses for the connected
contributions. The final estimates of the connected nucleon (proton) tensor charge 
for the isovector combination is 
$g_T^{u-d} = 1.020(76) $ in the $\overline{\text{MS}}$
scheme at $2\GeV$. 
The additional disconnected quark loop contributions needed for the
flavor-diagonal matrix elements are calculated using a stochastic
estimator employing the truncated solver method with the
all-mode-averaging technique.  We find that the size of the
disconnected contribution is smaller than the statistical error in the
connected contribution. This allows us to bound the disconnected
contribution and include it as an additional uncertainty in the
flavor-diagonal charges. After a continuum extrapolation, we find
$g_T^{u} = 0.774(66) $, $g_T^{d} = -0.233(28) $ and $g_T^{u+d} =
0.541(67) $. The strangeness tensor charge, that can make a significant
contribution to the neutron EDM due to the large ratio $m_s/m_{u,d}$,
is $g_T^{s}=0.008(9)$ in the continuum limit.
\end{abstract}
\maketitle
%
%
%
%
\section{Introduction}
\label{sec:into}

Precise estimates of the matrix elements of the isoscalar and
isovector tensor bilinear quark operators are needed to obtain bounds
on new physics from precision measurements of $\beta$-decays and
limits on the neutron electric dipole moment (nEDM). The isovector
charge, $g_T^{u-d}$, is needed to probe novel tensor interactions in
the helicity-flip part of the neutron decay
distribution~\cite{Bhattacharya:2011qm} while the isoscalar charges
are needed to quantify the contribution of the quark EDM to the nEDM
and set bounds on new sources of CP violation.  In this paper, we give
details of the simulations of lattice QCD using the clover-on-HISQ
approach to provide first principle estimates of these matrix elements
with control over all sources of systematic errors.

Lattice QCD analysis of isovector charges of nucleons is
well-developed (See the recent
reviews~\cite{Lin:2012ev,Syritsyn:2014saa,Constantinou:2014tga}).  In
this work we present precise estimates of this dominant contribution,
given by the connected diagrams, to the tensor charges, $i.e.$, the
insertion of the zero-momentum tensor operator in one of the three
valence quarks in the nucleon.  Calculation of the isoscalar charges
is similar except that it gets additional contributions from
contractions of the operator as a vacuum quark loop. This is called
the disconnected contribution as the quark loop and nucleon propagator
interact only through the exchange of gluons. The statistical signal
in the disconnected term is weak, so it is computationally much more
expensive. We find, on the four ensembles analyzed, that the
disconnected contributions of light quarks is small and in most cases
are consistent with zero within errors. We, therefore, use the largest of
these estimates to bound the disconnected contribution and include it
as a systematic uncertainty in the presentation of the final
results. Similarly, using five ensembles we show that the disconnected
contribution of the strange quark, also needed for the nEDM analysis,
is even smaller but we are able to extract a continuum limit
estimate~\cite{Bhattacharya:2015esa}.

Throughout the paper, we present results for the tensor charges of the
proton, which by convention are called nucleon tensor charges in
literature.  Results for the neutron are obtained by the $u
\leftrightarrow d$ interchange.  This paper is organized as
follows. In Section~\ref{sec:two}, we describe the parameters of the
gauge ensembles analyzed, the lattice methodology, fits used to
extract matrix elements within the ground state of the nucleon and the
renormalization of the operators. We discuss the calculation of the
connected diagram in section~\ref{sec:connected}, and of the
disconnected contribution in Section~\ref{sec:disc}.  Our final
results are presented in Section~\ref{sec:final_results} and we end
with conclusions in Section~\ref{sec:conclusions}.  In the Appendix we
present a summary of the control over systematics of existing
lattice calculations using the FLAG quality criteria~\cite{FLAGqc}.

\section{Lattice Parameters and Methodology}
\label{sec:two}

In this section we provide an overview of the calculational details.
These include a description of the gauge ensembles analyzed, a short
review of the operators used to calculate the two-point and
three-point correlation functions using clover fermions, the fit
ansatz used to extract the desired matrix elements from the
correlation functions and estimates of renormalization constants using
the RI-sMOM scheme.
%
\subsection{Lattice Parameters}
\label{sec:lat}

In order to obtain estimates with a desired precision, it is important
to quantify all sources of systematic errors. For matrix elements
between nucleon ground states, these include excited state
contamination, finite lattice volume, operator renormalization,
discretization effects at finite lattice spacing and extrapolations
from heavier $u$ and $d$ quarks.  Since lattice generation is very
expensive, it was, therefore, expedient to use a set of existing gauge
ensembles that cover a sufficiently large range in lattice spacing and
light quark mass to study the continuum and chiral behavior. The only
set available to us that meets our requirements are the ensembles
generated using $N_f=2+1+1$ flavors of highly improved staggered
quarks (HISQ)~\cite{Follana:2006rc} by the MILC
collaboration~\cite{Bazavov:2012xda}.  The parameters of the nine
ensembles used in this study are given in Table~\ref{tab:ens}.  In
this paper we show that these ensembles allow us to address issues of
statistics, excited state contamination, lattice volume, lattice
spacing and the chiral behavior in the calculation of the tensor
charges.

Staggered fermions have the advantage of being computationally cheaper
and preserve an important remnant of the continuum chiral
symmetry. Their disadvantage is that the spectrum has a four-fold
doubling in the continuum limit.  This doubling symmetry (called the
taste symmetry) is broken at finite lattice spacing and this breaking
introduces additional lattice artifacts. Due to taste mixing,
staggered baryon interpolating operators couple, in general, to a
combination of octet (the nucleon) and the decuplet (Delta)
states. Furthermore, these baryon operators couple to both parity
states in addition to all radial excitations of these.  Thus, baryon
correlation functions are more complicated to analyze compared to
Wilson-type fermions, as they have a weaker statistical signal, the
consequences of taste mixing has to be resolved and one has to take
into account the oscillating signal due to contributions from both
parity states. Since having a good statistical signal is a
prerequisite to quantifying the various sources of systematic errors,
we have chosen to construct correlation functions using Wilson-clover
fermions, as these preserve the continuum spin structure. This
mixed-action, clover-on-HISQ, approach, however, leads to a
nonunitary lattice formulation and at small, but {\it a priori}
unknown, quark masses suffers from the problem of exceptional
configurations discussed next.

Exceptional configurations are ones in which the clover Dirac operator
evaluated on HISQ configurations has near zero modes. As a result, on
such configurations the inversion of the clover Dirac operator, which
gives the Feynman propagator, can fail to converge and$/$or the
corresponding correlation functions have an exceptionally large
amplitude depending on the proximity to the zero mode.  The presence
of exceptional configurations biases the results or gives rise to
unphysically large fluctuations and invalidates the results.  In any
lattice analysis based on a unitary formulation, such as HISQ-on-HISQ
or clover-on-clover, such configurations are suppressed in the lattice
generation. Given an appropriately generated ensemble of HISQ
configurations, there is no basis for excluding any configuration from
the ensemble average in a clover-on-HISQ calculation.  Thus, these 
calculations should be done only on ensembles without any exceptional
configurations.

The presence of such exceptional configurations in a clover-on-HISQ
analysis is expected to increase on decreasing the quark mass at fixed
lattice spacing and increase with the lattice spacing at fixed quark
mass, $i.e.$, the coarser the configurations, the more likely they
are. Consequently, smearing techniques used to reduce short distance
lattice artifacts also reduce the probability of encountering
exceptional configurations. To reduce lattice artifacts, we applied
HYP smearing~\cite{Hasenfratz:2001hp} to all HISQ configurations.  To
check for exceptional configurations on these HYP smeared lattices, we
monitor the convergence of the quark propagator and the size of
fluctuations in correlations functions on each configuration.  These
tests provided evidence for exceptional configurations on the
$a=0.15$~fm ensembles and on the $a=0.12$~fm ensemble with $M_\pi \approx
130$~MeV. Consequently, these ensembles, also generated by the MILC
collaboration, were excluded from our analysis.  An earlier discussion
regarding exceptional configurations on these ensembles is given in
Ref.~\cite{Bhattacharya:2013ehc}.  To reiterate, the nine ensembles
used in this study and described in Table~\ref{tab:ens} did not
present evidence of an exceptional configuration.

\begin{table}
\begin{center}
\renewcommand{\arraystretch}{1.2} 
\begin{ruledtabular}
\begin{tabular}{lc<{ }|cccc}
\multicolumn{2}{c}{Ensemble ID} & $a$ (fm) & $M_\pi^{\rm sea}$ (MeV) & $L^3\times T$    & $M_\pi L$
\\ \hline
a12m310 & \includegraphics[viewport=14 13 20 19]{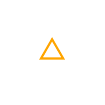}   & 0.1207(11) & 305.3(4) & $24^3\times 64$ & 4.55 \\
a12m220S& \includegraphics[viewport=14 13 20 19]{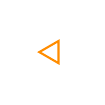}  & 0.1202(12) & 218.1(4) & $24^3\times 64$ & 3.29 \\
a12m220 & \includegraphics[viewport=14 13 20 19]{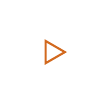}   & 0.1184(10) & 216.9(2) & $32^3\times 64$ & 4.38 \\
a12m220L& \includegraphics[viewport=14 13 20 19]{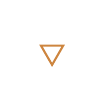}  & 0.1189(09) & 217.0(2) & $40^3\times 64$ & 5.49 \\
\hline
a09m310 & \includegraphics[viewport=14 13 20 19]{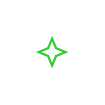}   & 0.0888(08) & 312.7(6) & $32^3\times 96$ & 4.51 \\
a09m220 & \includegraphics[viewport=14 13 20 19]{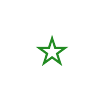}   & 0.0872(07) & 220.3(2) & $48^3\times 96$ & 4.79 \\
a09m130 & \includegraphics[viewport=14 13 20 19]{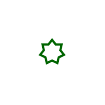}   & 0.0871(06) & 128.2(1) & $64^3\times 96$ & 3.90 \\
\hline
a06m310 & \includegraphics[viewport=14 13 20 19]{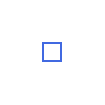}   & 0.0582(04) & 319.3(5) & $48^3\times 144$& 4.52 \\
a06m220 & \includegraphics[viewport=14 13 20 19]{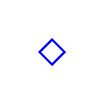}   & 0.0578(04) & 229.2(4) & $64^3\times 144$& 4.41 \\
\end{tabular}
\end{ruledtabular}
\caption{Parameters of the (2+1+1) flavor HISQ lattices generated by the MILC 
collaboration and analyzed in this study are quoted from Ref.~\cite{Bazavov:2012xda}. 
Symbols used in plots are defined along with the ensemble ID. 
Finite size effects are analyzed in terms of $M_\pi L$ with the 
clover-on-HISQ $M_\pi$ defined in Table~\ref{tab:cloverparams}.
}
\label{tab:ens}
\end{center}
\end{table}

The parameters used in the analysis with clover fermions are given in
Table~\ref{tab:cloverparams}. The Sheikholeslami-Wohlert coefficient used
in the clover action is fixed to its tree-level value with tadpole
improvement, $c_\text{sw} = 1/u_0^3$ where $u_0$ is the tadpole factor
of the HYP smeared HISQ lattices.

The masses of light clover quarks were tuned so that the
clover-on-HISQ pion masses $M_\pi$ (see Table~\ref{tab:cloverparams})
match the HISQ-on-HISQ Goldstone ones, $M_\pi^{\rm sea}$, given in
Table~\ref{tab:ens}. The strange quark mass $m_s$ is tuned so that the
resulting clover-on-HISQ $M_{s\bar{s}} = \sqrt{m_s^b/m_l^b}
M_\pi^\text{sea}$, where $m_s^b/m_l^b$ is the ratio of bare strange
and light quark masses used in the HISQ generation, and is 5 for m310
lattices, 10 for m220 lattices and 27 for m130 lattices.  The
resulting estimates for $m_s$ are consistent with those obtained by
matching to the kaon mass~\cite{Briceno:2012wt}.

All fits in $M_\pi^2$ to study the chiral
behavior are made using the clover-on-HISQ $M_\pi^2$ as correlation functions
and thus the observables have a greater sensitivity to it. Performing
fits using the HISQ-on-HISQ values of ${M_\pi^{\rm sea}}^2$ did not change the
estimates significantly.

Estimates of nucleon charges and form-factors based on lower statistics subset of 
data presented here for the two ensembles {\it a12m310} and 
{\it a12m220} have been published in~\cite{Bhattacharya:2013ehc}. Results
for the tensor charges presented in this paper supersede those earlier
estimates.

\begin{table*}
\centering
\begin{ruledtabular}
\begin{tabular}{c|ccccl}
ID       & $m_l$ & $m_s$ & $c_{\text{SW}}$ & $M_\pi^{\text{val}}~(\MeV)$ & Smearing  \\
\hline
a12m310  & $-$0.0695  & $-$0.018718& 1.05094 & 310.2(2.8) & \{5.5, 70\} [\{5.5, 70\}] \\
a12m220S & $-$0.075   & ---        & 1.05091 & 225.0(2.3) & \{5.5, 70\} \\
a12m220  & $-$0.075   & $-$0.02118 & 1.05091 & 227.9(1.9) & \{5.5, 70\} [\{5.5, 70\}] \\
a12m220L & $-$0.075   & ---        & 1.05091 & 227.6(1.7) & \{5.5, 70\} \\ \hline
a09m310  & $-$0.05138 & $-$0.016075& 1.04243 & 313.0(2.8) & \{5.5, 70\} [\{6.0, 80\}] \\
a09m220  & $-$0.0554  & $-$0.01761 & 1.04239 & 225.9(1.8) & \{5.5, 70\} [\{6.0, 80\}] \\
a09m130  & $-$0.058   & ---        & 1.04239 & 138.1(1.0) & \{5.5, 70\} \\ \hline
a06m310  & $-$0.0398  & $-$0.01841 & 1.03493 & 319.6(2.2) & \{6.5, 70\} [\{6.5, 80\}] \\
a06m220  & $-$0.04222 & ---        & 1.03493 & 235.2(1.7) & \{5.5, 70\} \\
\end{tabular}
\end{ruledtabular}
\caption{The parameters used in the calculation of clover propagators.
  The hopping parameter $\kappa$ in the clover action is given by
  $2\kappa_{l,s} = 1/(m_{l,s}+4)$.  $m_s$ is needed for the
  calculation of the strange quark disconnected loop diagram. The
  Gaussian smearing parameters are defined by $\{\sigma,
  N_{\text{KG}}\}$ where $N_{\text{KG}}$ is the number of applications
  of the Klein-Gordon operator and the width is controlled by the
  coefficient $\sigma$, in Chroma convention.  Smearing parameters
  used in the study of disconnected diagrams are given within
  square-brackets.  $m_l$ is tuned to achieve $M_\pi =
  M_\pi^\text{sea}$, and $m_s$ is tuned so that $M_{s\bar{s}} =
  \sqrt{m_s^b/m_l^b} M_\pi^\text{sea}$.  The error in the pion mass $M_\pi$ is governed
  mainly by the uncertainty in the lattice scale given in
  Table~\protect\ref{tab:ens}.}
  \label{tab:cloverparams}
\end{table*}

\subsection{Lattice Methodology}
\label{sec:methodology}
The two-point and three-point nucleon correlation functions at zero momentum 
are defined as 
\begin{align}
 C_{\alpha \beta}^{\text{2pt}}(t)
  &= \sum_{\mathbf{x}} 
   \langle 0 \vert \chi_\alpha(t, \mathbf{x}) \overline{\chi}_\beta(0, \mathbf{0}) 
   \vert 0 \rangle \,, \\
 C_{\Gamma; \alpha \beta}^{\text{3pt}}(t, \tau)
  &= \sum_{\mathbf{x}, \mathbf{x'}} 
  \langle 0 \vert \chi_\alpha(t, \mathbf{x}) \mathcal{O}_\Gamma(\tau, \mathbf{x'})
  \overline{\chi}_\beta(0, \mathbf{0}) 
   \vert 0 \rangle \,,
\label{eq:corr_funs}
\end{align}
where $\alpha$ and $\beta$ are the spinor indices. The source
time slice is translated to $t_0=0$, $t$ is the sink time slice, and $\tau$ is
the time slice at which the bilinear operator $\mathcal{O}_\Gamma^q(x)
= \bar{q}(x) \Gamma q(x)$ is inserted. The Dirac matrix $\Gamma$ is
$1$, $\gamma_4$, $\gamma_i \gamma_5$ and $\gamma_i \gamma_j$ for
scalar (S), vector (V), axial (A) and tensor (T) operators,
respectively.
In this paper, subscripts $i$ and $j$ on gamma matrices run over $\{1,2,3\}$, 
with $i<j$. 
The interpolating operator used to create$/$annihilate the nucleon
state, $\chi$, is 
\begin{align}
 \chi(x) = \epsilon^{abc} \left[ {q_1^a}^T(x) C \gamma_5 
            \frac{1}{2} (1 \pm \gamma_4) q_2^b(x) \right] q_1^c(x)
\label{eq:nucl_op}
\end{align}
with color indices $\{a, b, c\}$, charge conjugation matrix $C$, and
$q_1$ and $q_2$ denoting the two different flavors of light quarks.
The nonrelativistic projection $(1 \pm \gamma_4)/2$ is inserted to
improve the signal, with the plus and minus sign applied to the
forward ($t>0$) and backward ($t<0$) propagation, respectively.

The nucleon charges $g_\Gamma^q$ are  defined as 
\begin{align}
 \langle N(p, s) \vert \mathcal{O}_\Gamma^q \vert N(p, s) \rangle
 = g_\Gamma^q \bar{u}_s(p) \Gamma u_s(p)
\end{align}
with spinors satisfying
\begin{align}
 \sum_s u_s(\mathbf{p}) \bar{u}_s(\mathbf{p})  = {\qslash{p} + m_N} \,.
\end{align}
These charges, $g_\Gamma^q$, can be extracted from the ratio of the
projected three-point function to the two-point function for $t \gg
\tau \gg 0$
\begin{align}
 R_\Gamma(t, \tau) 
  & \equiv \frac{\langle \Tr [ \mathcal{P}_\Gamma C_{\Gamma}^{\text{3pt}}(t, \tau) ]\rangle}
         {\langle \Tr [ \mathcal{P}_\text{2pt} C^{\text{2pt}}(t) ] \rangle} \nonumber \\
  & \longrightarrow \frac{1}{8} \Tr 
   \left[\mathcal{P}_\Gamma (1+\gamma_4) \Gamma (1+\gamma_4) \right] 
   g_\Gamma^q.
 \label{eq:3pt_2pt_ratio}
\end{align}
Here $\mathcal{P}_\text{2pt} = (1+\gamma_4)/2$ is used to project out
the positive parity contribution and $\mathcal{P}_\Gamma$ is defined
below. Note that the ratio in Eq.~\eqref{eq:3pt_2pt_ratio} becomes
zero if $\Gamma$ anti-commutes with $\gamma_4$, so only $\Gamma = 1$,
$\gamma_4$, $\gamma_i \gamma_5$ and $\gamma_i \gamma_j$ can survive.
In this paper, we present results for only the tensor channel, for
which we can demonstrate control over all systematic errors.

\begin{figure*}[tb]
  \subfigure{
    \includegraphics[width=0.4\linewidth]{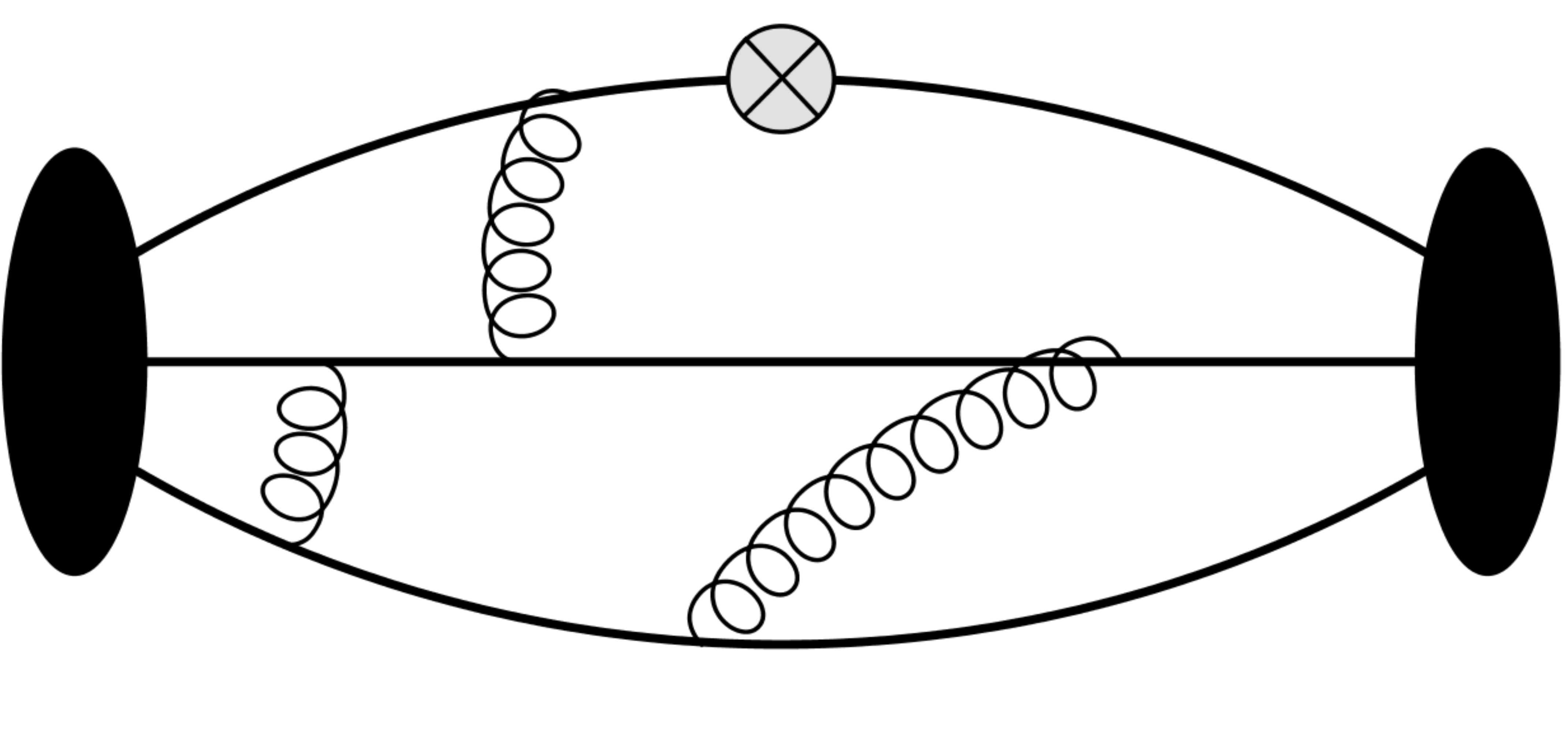}
  }
  \hspace{0.04\linewidth}
  \subfigure{
    \includegraphics[width=0.4\linewidth]{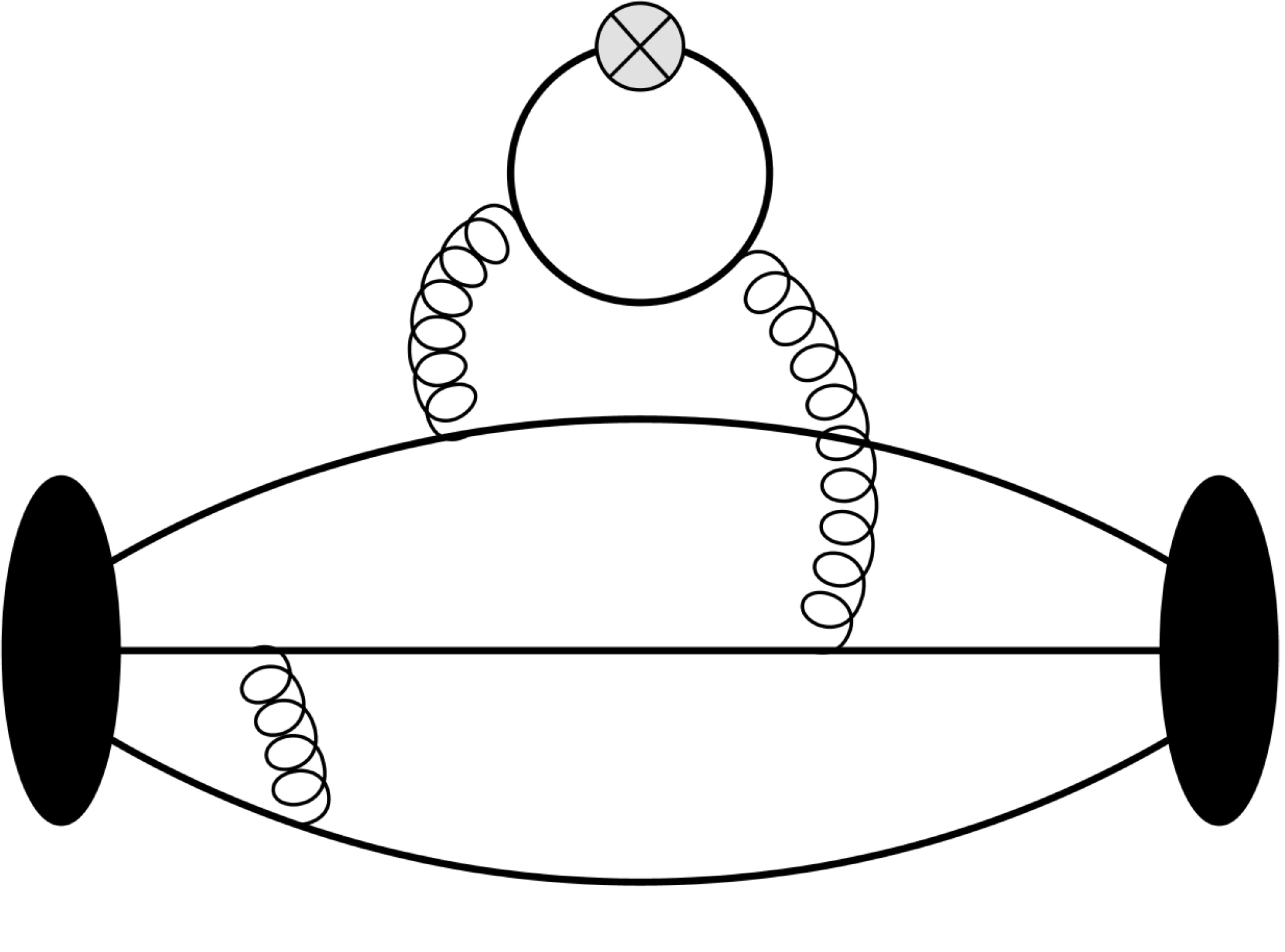}
  }
\caption{The connected (left) and disconnected (right)  three-point diagrams needed 
  to calculate the matrix elements of bilinear quark operators in the nucleon state.
  \label{fig:con_disc}}
\end{figure*}

On inserting a bilinear quark operator between the nucleon states to
construct the three-point function, one gets two classes of diagrams:
(i) the bilinear operator is contracted with one of the three valence
quarks in the nucleon, as shown in the left diagram of
Fig.~\ref{fig:con_disc}, and (ii) the bilinear operator is contracted
into a quark loop that is correlated with the nucleon two-point
function through the exchange of gluons, as shown in the right diagram
of Fig.~\ref{fig:con_disc}.  These are called the connected and
disconnected diagrams, respectively.

The disconnected part of Eq.~\eqref{eq:3pt_2pt_ratio} can be written as
\begin{align}
 R_\Gamma^\text{disc}&(t, \tau)
   =  \langle \sum_\mathbf{x} \Tr [M^{-1}(\tau, \mathbf{x}; \tau, \mathbf{x}) \Gamma] 
     \rangle \nonumber \\
  &- \frac{\langle \Tr [ \mathcal{P}_\Gamma C^{\text{2pt}}(t) ] 
    ~\sum_\mathbf{x} \Tr [M^{-1}(\tau, \mathbf{x}; \tau, \mathbf{x}) \Gamma] 
    \rangle}
   {\langle \Tr [ \mathcal{P}_\text{2pt} C^{\text{2pt}}(t) ] \rangle}
 \,,
 \label{eq:3pt_2pt_ratio_disc}
\end{align}
where $M$ is the Dirac operator.
Note that the first term of the right-hand side is zero when $\Gamma \neq 1$,
so it does not contribute to the tensor charges. High precision measurements 
of Eq.~\eqref{eq:3pt_2pt_ratio_disc} requires improving the signal in the
second term, $i.e.$, the correlation between the nucleon two-point functions 
$C^{\text{2pt}}$ and the quark loop $\Tr[M^{-1}\Gamma]$ as discussed in Sections~\ref{sec:2ptD} 
and~\ref{sec:DQLC}. 

The charges $g_\Gamma^q$ are extracted from the ratio $R_\Gamma$ by 
appropriate choice of the projection operator $\mathcal{P}_\Gamma$. 
For the calculation of connected contribution we use a single
  projection operator $\mathcal{P}_\Gamma =
  \mathcal{P}_\text{2pt}(1+i\gamma_5\gamma_3)$ to extract all four
  tensor structures at the same time as the projection is done at the
  time of the calculation of the sequential propagator.  For the
  disconnected diagram, the projection operator is part of the final
  trace with the two-point function, so there is no additional cost to
  using tensor specific projectors. We, therefore, use
\begin{align}
 \begin{split}
 \mathcal{P}_1 &= \mathcal{P}_\text{2pt}, \\
 \mathcal{P}_{\gamma_i \gamma_5} &= \mathcal{P}_\text{2pt} \gamma_5 \gamma_i \quad (i=1,2,3), \\
 \mathcal{P}_{\gamma_1 \gamma_2} &= \mathcal{P}_\text{2pt} \gamma_5 \gamma_3, \\
 \mathcal{P}_{\gamma_2 \gamma_3} &= \mathcal{P}_\text{2pt} \gamma_5 \gamma_1, \\
 \mathcal{P}_{\gamma_1 \gamma_3} &= \mathcal{P}_\text{2pt} \gamma_2 \gamma_5,
 \end{split}
 \label{eq:projs}
\end{align}
which make $R_\Gamma(t,\tau) \rightarrow g_\Gamma$ for $t \gg \tau \gg 0$ as they satisfy
\begin{align}
\frac{1}{8} \Tr \left[ \mathcal{P}_\Gamma (1+\gamma_4) \Gamma (1+\gamma_4) \right] = 1 \,.
\end{align}

For the disconnected diagrams, the statistics in the two-point function are
improved by including the backward propagating baryons ($t < 0$)
from each source point. In this case $\mathcal{P}^\text{2pt} =
(1-\gamma_4)/2$ is used to project out the negative parity state and
we multiply the tensor projection operators
$\mathcal{P}_{\gamma_{i}\gamma_{j}}$ in Eq.~\eqref{eq:projs} by $(-1)$
to match the convention used for forward propagation.

The calculation of the two-point and the connected three-point
functions is carried out using the method described in our previous
study, Ref.~\cite{Bhattacharya:2013ehc}. These calculations use the
Chroma software package \cite{Edwards:2004sx}. Part of the
calculations are done on clusters with graphic processing units (GPUs)
using QUDA library \cite{Clark:2009wm}.  The source and sink baryon
operators are constructed using smeared quark propagators to reduce
the contamination from the excited states.  We use gauge-invariant
Gaussian smeared sources to improve the overlap with the ground
state. Smearing is done by applying the three-dimensional Klein-Gordon
operator $\nabla^2$ a fixed number of times $(1 - \sigma^2\nabla^2/(4N_{\rm KG}))^{N_{\rm KG}}$.  The smearing parameters
$\{\sigma, N_{\rm KG}\}$ for each ensemble are given in
Table~\ref{tab:cloverparams}.

To calculate the connected three-point function, we analyze
configurations in sets of four measurements, $i.e.$ we generate four
independent propagators $S_s^o$ on each configuration using smeared
sources on four maximally separated time slices $t_{\rm src}^{i=1,4}$.
For each $S_s^o$, the same smearing operation is applied at all
time slices to create the smeared sink, and the two-point correlation
function is calculated using these smeared-smeared propagators
$S_{ss}$.  Each of these four smeared propagators are used to
construct sources for $u$ and $d$ quark propagators with the insertion
of zero-momentum nucleon state at time slices displaced by a fixed
$t_{\rm sep}$ from the four source time slices $t_{\rm src}^{i=1,4}$.
These $u$ and $d$ sequential sources (generated separately) at $t_{\rm
  src}^{i=1,4}+ t_{\rm sep}$ are smeared again.  The final coherent
sequential propagator $S_c^{seq}$ is then calculated using the sum of
these four smeared sources, $i.e.$, the sequential propagator from the
four time slices is calculated at one go. The connected three-point
functions, over the four regions $t_{\rm src}^{i=1,4}$ to $t_{\rm
  src}^{i=1,4}+ t_{\rm sep}$, are then constructed by inserting the
bilinear operator between each of the original individual propagator
$S_s^o$ from $t_{\rm src}^{i=1,4}$ and the coherent sequential
propagator $S_c^{seq}$ from $t_{\rm src}^{i=1,4}+ t_{\rm
  sep}$~\cite{Yamazaki:2009zq}. The assumption one makes by adding the
four sources to produce a single coherent sequential propagator is
that the entire contribution to the three-point function in any one of
the four intervals between $t_{\rm src}^{i=1,4}$ and $t_{\rm
  src}^{i=1,4}+ t_{\rm sep}$ is from baryon insertion at $t_{\rm
  src}^{i}+ t_{\rm sep}$ and the contribution of baryon sources at the
other three time slices $t_{\rm src}^{j \neq i}+ t_{\rm sep}$ in
$S_c^{seq}$ goes to zero on averaging over the gauge configurations.
The coherent sequential source method has the advantage that
the insertion of operators with different tensor structures and
various momenta and for all four source positions can be done at the
same time with tiny computational overhead.

To study and quantify the excited state contamination, we repeat the
calculation for multiple source-sink separations, $t_\text{sep}$.
Separate sequential $u$ and $d$ propagators are calculated for each
$t_{\rm sep}$ analyzed. Thus, the total number of inversions of the
Dirac operator are $4 + 2 \times N_{t_\text{sep}}$ for each set of
four measurements on each configuration. Our choices of $t_\text{sep}$
and the number of measurements made on each ensemble (number of
configurations times the number of sources on each configuration) are
given in Table~\ref{tab:tsep}.

The calculation of disconnected quark loop diagrams using stochastic
methods have a poor signal and requires very high statistics. Because
of the computational cost, the calculations with light quarks have
been done on the three heaviest, $M_\pi \approx 310$~MeV, and the 
{\it a12m220} ensembles; and on five ensembles for the strange quark as
listed in Table~\ref{tab:params_disc}.
For the evaluation of the disconnected diagrams, we obtain a
stochastic estimate using the truncated solver method (TSM)
\cite{Collins:2007mh, Bali:2009hu} with the all-mode-averaging (AMA)
technique \cite{Blum:2012uh} as described in Section~\ref{sec:disc}.

%

\subsection{Fits to Correlation Functions}
\label{sec:fits}

To extract the desired nucleon charges, the matrix elements of the
bilinear quark operators need to be calculated between ground state
nucleons.
On the lattice, however, any zero-momentum correlation function
defined in~Eq.~\eqref{eq:corr_funs} using the nucleon interpolation
operator defined in~\eqref{eq:nucl_op}, has a coupling to the ground
state nucleon, all radially excited states, and multiparticle states
with the same quantum numbers.  Operators constructed using
appropriately tuned smeared sources reduce the coupling to excited
states but do not eliminate it. We discuss two synergistic strategies
for removing the remaining excited state contamination based on the
fact that in Euclidean time, the contributions from the excited states
are exponentially suppressed as (i) the distance between the
source/sink and the inserted operator increases and (ii) as the
mass gap, $M_{\rm excited}-M_0$, increases. First, one can increase
$t_{\rm sep}$.  Unfortunately, the signal also decreases exponentially
as $t_{\rm sep}$ is increased so one is forced to compromise.  In
fact, the values of $t_{\rm sep}$ we have used reflect this compromise
based on the anticipated statistics on each ensemble.  Additionally,
one can include excited states in the analysis of
Eq.~\eqref{eq:corr_funs} as discussed below.

We include one excited state in the analysis of the two- and
three-point functions. For operator insertion at zero momentum, the
data are fit using the ansatz\"e
\begin{align}
C^\text{2pt}
  &(t_f,t_i) = \nonumber \\
  &{|{\cal A}_0|}^2 e^{-M_0 (t_f-t_i)} + {|{\cal A}_1|}^2 e^{-M_1 (t_f-t_i)}\,,\\
C^\text{3pt}_{\Gamma}&(t_f,\tau,t_i) = \nonumber\\
  & |{\cal A}_0|^2 \langle 0 | \mathcal{O}_\Gamma | 0 \rangle  e^{-M_0 (t_f-t_i)} +{}\nonumber\\
  & |{\cal A}_1|^2 \langle 1 | \mathcal{O}_\Gamma | 1 \rangle  e^{-M_1 (t_f-t_i)} +{}\nonumber\\
  & {\cal A}_0{\cal A}_1^* \langle 0 | \mathcal{O}_\Gamma | 1 \rangle  e^{-M_0 (\tau-t_i)} e^{-M_1 (t_f-\tau)} +{}\nonumber\\
  & {\cal A}_0^*{\cal A}_1 \langle 1 | \mathcal{O}_\Gamma | 0 \rangle  e^{-M_1 (\tau-t_i)} e^{-M_0 (t_f-\tau)} ,
\label{eq:2pt_3pt}
\end{align}
where the source positions are shifted to $t_i=0$ and $t_f = t_\text{sep}$. 
The states $|0\rangle$ and $|1\rangle$
represent the ground and ``first'' excited nucleon states,
respectively. The four parameters, $M_0$, $M_1$, ${\cal A}_0$ and
${\cal A}_1$ are estimated first from the two-point function data.  We find
that the extraction of $M_0$ and ${\cal A}_0$ is stable under change
of the fit range, while that of $M_1$ and ${\cal A}_1$ is not.  We,
therefore, choose the largest range, requiring that the values of, and
the errors in, all four parameters do not jump by a large amount on
changing the fit range.  In all these fits, we find $M_1 \approx 2
M_0$, so it should be considered an effective excited state mass as it
is much larger than the $N(1440)$ excitation.  The results of these best
fits are given in Table~\ref{tab:res2pt}.

We performed two independent measurements of the two-point functions,
and the corresponding $M_0$, $M_1$, ${\cal A}_0$ and ${\cal A}_1$ are
given in Table~\ref{tab:res2pt}.  The second set of measurements were
obtained during the calculation of the disconnected diagrams using the
AMA error reduction method discussed in Section~\ref{sec:2ptD}.  We
find that the two estimates are consistent within errors indicating no
remaining bias with our choice of parameters for the AMA.

\begin{figure*}[tb]
  \subfigure{
    \includegraphics[width=0.33\linewidth]{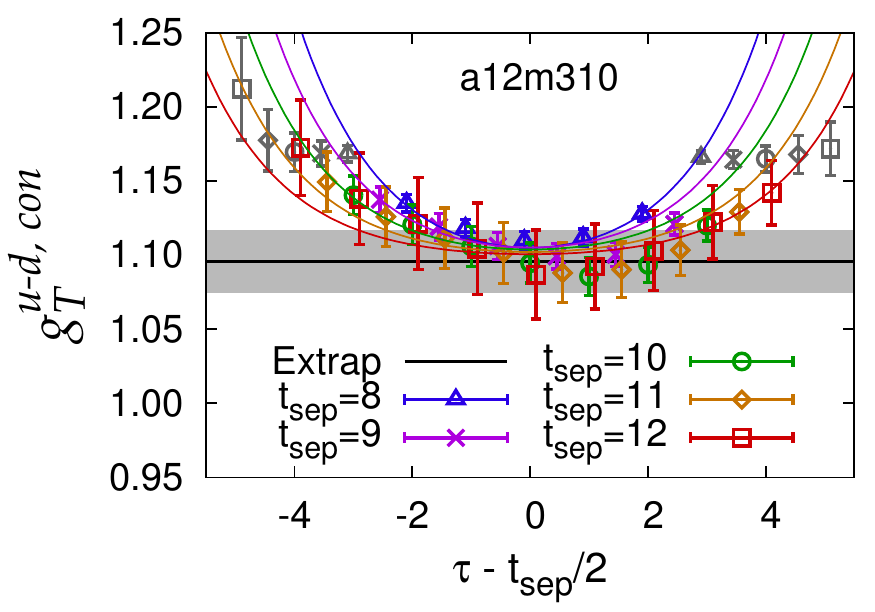}
    \includegraphics[width=0.33\linewidth]{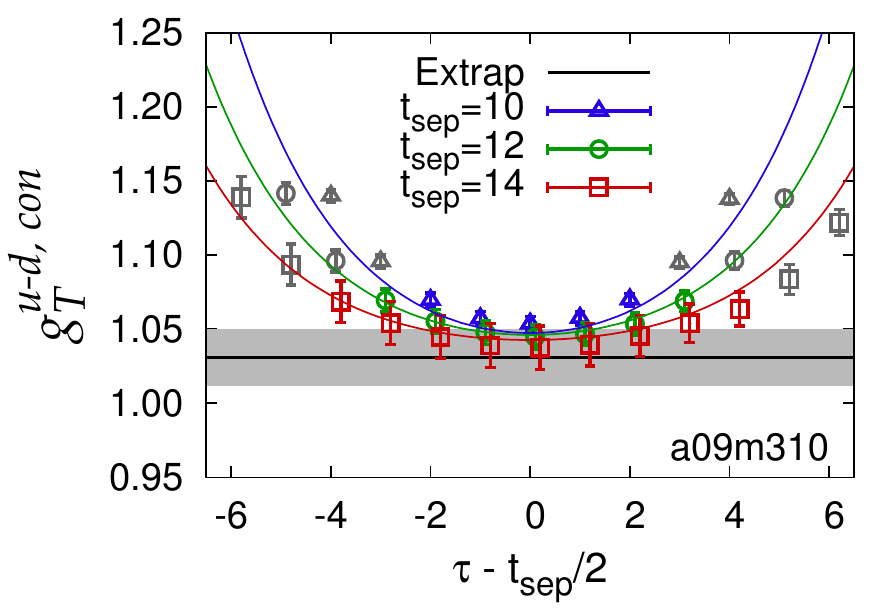}
    \includegraphics[width=0.33\linewidth]{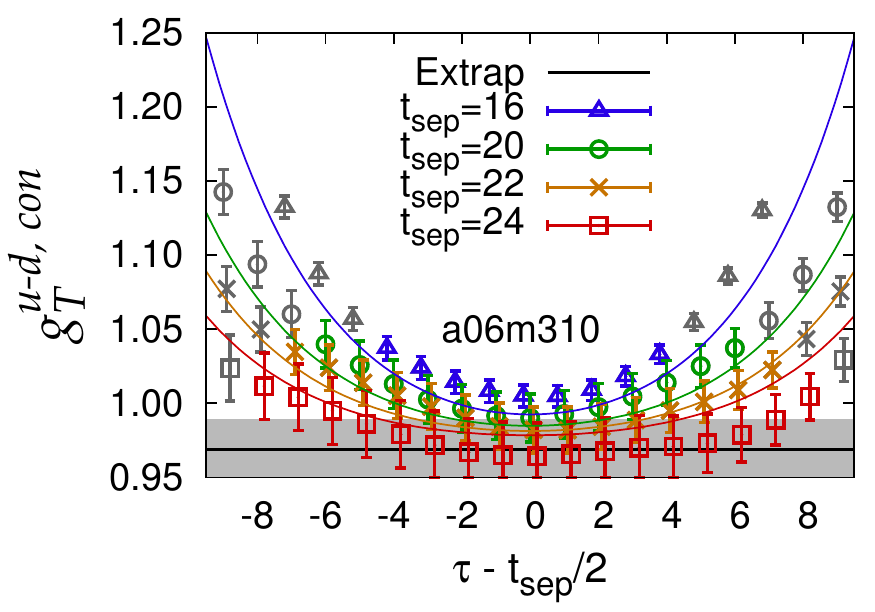}
  }
  \hspace{0.04\linewidth}
  \subfigure{
    \includegraphics[width=0.33\linewidth]{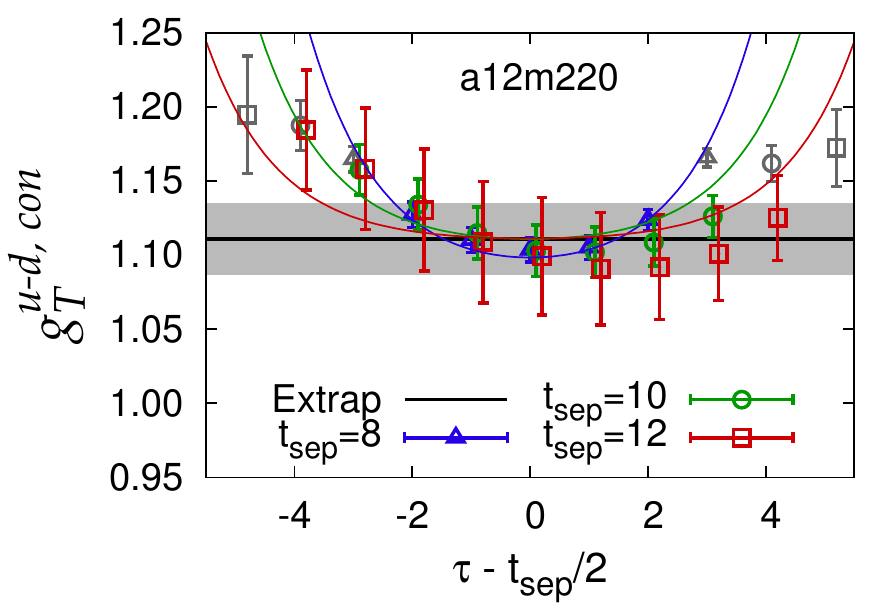}
    \includegraphics[width=0.33\linewidth]{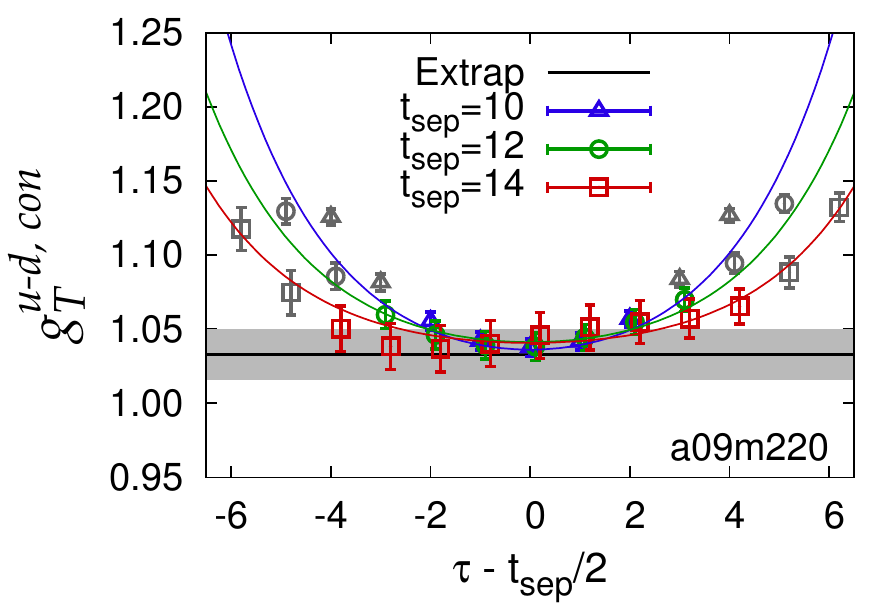}
    \includegraphics[width=0.33\linewidth]{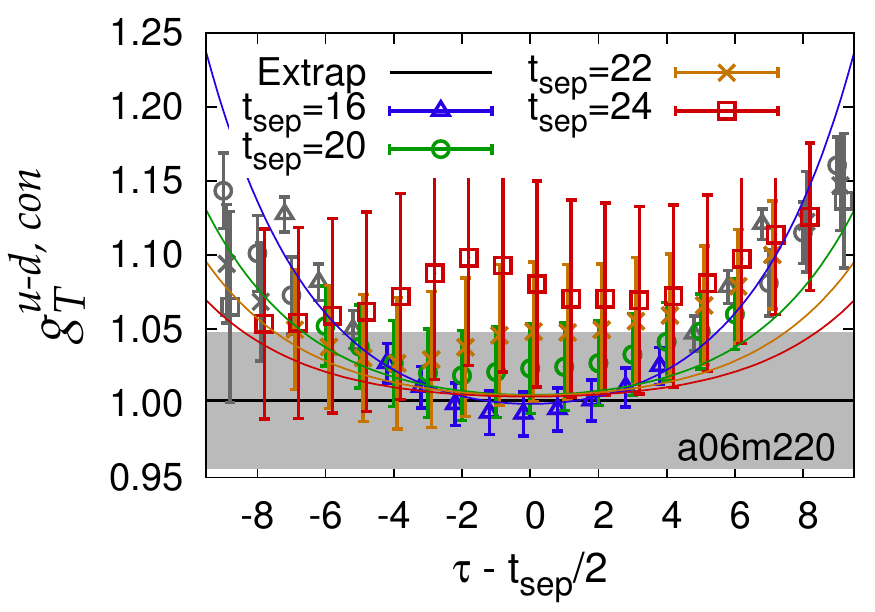}
  }
  \hspace{0.04\linewidth}
  \subfigure{
    \includegraphics[width=0.32\linewidth]{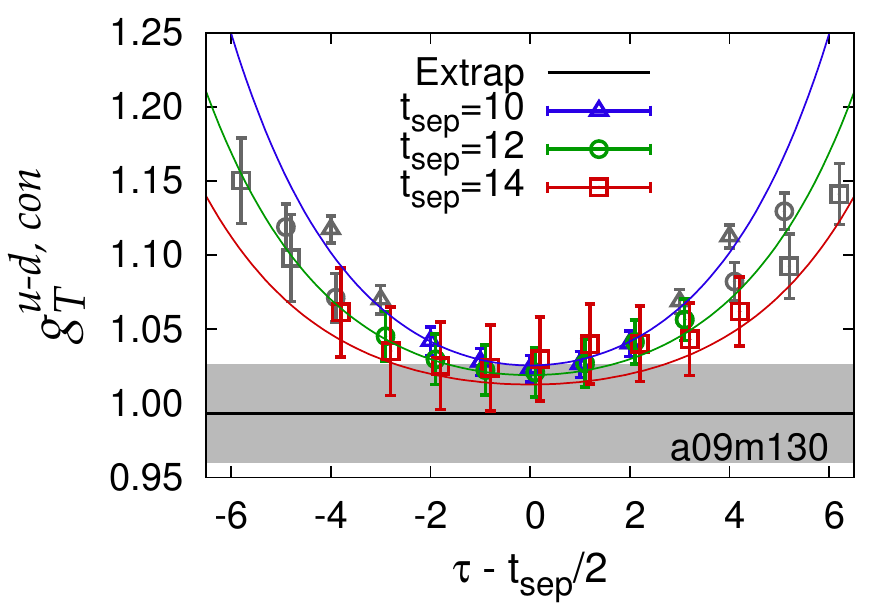}
  }
\caption{The data for $g_T^{u-d}$ and the results of the simultaneous
  fit using multiple $t_{\rm sep}$ using the ansatz given in
  Eq.~\protect\eqref{eq:2pt_3pt} to isolate the excited state
  contribution. The seven figures are arranged as follows: the $M_\pi
  \approx 310$~MeV ensembles (top), $M_\pi \approx 220$~MeV ensembles (middle) and
  the $M_\pi \approx 130$~MeV ensemble (bottom).  The solid black line and
  the grey band are the ground state ($t_\text{sep} \to \infty$)
  estimate and error. The fits evaluated for different $t_\text{sep}$
  are also shown by solid lines.
  \label{fig:excited_conn}}
\end{figure*}

Fits using the ansatz for the three-point function given in
Eq.~\eqref{eq:2pt_3pt} are used to isolate the two unwanted matrix
elements $\langle 0 | \mathcal{O}_\Gamma | 1 \rangle$ and $\langle 1 |
\mathcal{O}_\Gamma | 1 \rangle$.  We find that the magnitude of
$\langle 0 | \mathcal{O}_\Gamma | 1 \rangle$ is about $16\%$ of
$\langle 0 | \mathcal{O}_\Gamma | 0 \rangle$ and is determined with
about $20\%$ uncertainty on all the ensembles, whereas $|\langle 1 |
\mathcal{O}_\Gamma | 1 \rangle| \sim \langle 0 | \mathcal{O}_\Gamma |
0 \rangle$, but has $O(100\%)$ errors. Ideally, equally precise data
should be generated at each value of $t_{\rm sep}$. In our analysis,
however, the same number of measurements have been made for all
$t_{\rm sep}$ on each ensemble, so errors increase with $t_{\rm sep}$
as shown in Fig.~\ref{fig:excited_conn}.

We reduce the contamination from higher excited states in these fits
to the two- and three-point functions by excluding data points
overlapping with, and adjacent to, the source and sink time slices at
which the excited state contamination is the largest. For uniformity,
we exclude $2,\ 3,\ 4$ time slices on either end of the interval
$t_{\rm src}$ to $t_{\rm src} + t_{\rm sep}$ in the three-point
function for the $a=0.12$, $a=0.09$ and $a=0.06$~fm ensembles,
respectively.  In physical units, these excluded regions correspond to
roughly the same distance.  This range of excluded points is
consistent with the starting time slice of the fits to the two-point
correlators given in Table~\ref{tab:res2pt}, $i.e.$, the time beyond
which a two-state fit captures the two-point function data. Changing
the exclusion time slice values to $3,\ 4,$ and $6$ in both the fits
changed the final estimates of the charges by less than $1\sigma$.

\begin{table}
\begin{center}
\begin{ruledtabular}
\begin{tabular}{c|ccc}
ID & $t_\text{sep}/a$ & $N_\text{conf}$ & $N_\text{meas}$  \\
\hline
a12m310 & $\{8,9,10,11,12\}$& 1013 & 8104  \\
a12m220S& $\{8, 10, 12\}$   & 1000 & 24000 \\
a12m220 & $\{8, 10, 12\}$   & 958  & 7664  \\
a12m220L& $10$              & 1010 & 8080  \\
\hline                             
a09m310 & $\{10,12,14\}$    & 881  & 7048  \\
a09m220 & $\{10,12,14\}$    & 890  & 7120  \\
a09m130 & $\{10,12,14\}$    & 883  & 7064  \\
\hline                             
a06m310 & $\{16,20,22,24\}$ & 1000 & 8000  \\
a06m220 & $\{16,20,22,24\}$ & 650  & 2600  \\
\end{tabular}
\end{ruledtabular}
\caption{The values of source-sink time separations ($t_\text{sep}/a$)
  used, the total number of configurations analyzed ($N_\text{conf}$)
  and measurements made ($N_\text{meas}$) for the two-point and
  connected three-point function calculations. }
\label{tab:tsep}
\end{center}
\end{table}

%

%

Including a second exited state in the analysis would increase the
number of matrix elements to be estimated from the three-point
function by three. Given that their contribution would be smaller
still, much higher statistics than generated for this study would be
needed.  This is confirmed in practice; the data are well fit with
just the one excited state ansatz and there is no sensitivity left to
resolve three additional small parameters.  This analysis, using the
ansatz\"e given in Eq.~\eqref{eq:2pt_3pt} and fitting the data at all
$t_{\rm sep}$ simultaneously, was called the {\it two-simRR} method in
Ref.~\cite{Bhattacharya:2013ehc}.

Our overall conclusion is that, using the values of $t_{\rm sep}$ and
the statistics for each ensemble given in Table~\ref{tab:tsep}, and
assuming that only one excited state gives a significant contribution,
we are able to isolate and remove this contamination as illustrated in
Fig.~\ref{fig:excited_conn}. It turns out that on all nine ensembles,
the excited state contamination for $g_T$ is small. It is worth
remarking that this is not the case for $g_A$ as discussed
in~\cite{Bhattacharya:2015gA}.

\begin{table*}
\centering
\begin{ruledtabular}
\begin{tabular}{c|ccccc|ccccc}
ID       & Fit Range & $aM_0$      & $aM_1$     & ${\cal A}_0 \times 10^{11}$  & ${\cal A}_1 \times 10^{11}$   
         & Fit Range & $aM_0$      & $aM_1$     & ${\cal A}_0 \times 10^{11}$  & ${\cal A}_1 \times 10^{11}$  \\
\hline
a12m310  & 2--15     & 0.6669(53) & 1.36(11)  & 6.57(27)  & 6.28(61)   & 2--15   & 0.6701(16) & 1.471(45) & 6.845(82) & 6.88(35)  \\
a12m220S & 2--15     & 0.6233(55) & 1.42(13)  & 6.58(26)  & 6.94(93)   &         &            &           &           &           \\
a12m220  & 2--15     & 0.6232(49) & 1.45(15)  & 6.58(24)  & 6.8(1.1)   & 2--15   & 0.6124(17) & 1.294(37) & 6.070(91) & 6.34(23)  \\
a12m220L & 2--15     & 0.6046(71) & 1.16(12)  & 5.68(37)  & 5.63(51)   &         &            &           &           &           \\
\hline                                                                                                                
a09m310  & 3--20     & 0.4965(46) & 0.938(57) & 14.12(75) & 17.4(1.1)  & 3--20   & 0.4973(12) & 0.971(22) & 2.215(31) & 2.374(74) \\
a09m220  & 3--20     & 0.4554(45) & 0.925(53) & 12.13(61) & 18.5(1.3)  & 3--20   & 0.4524(24) & 0.877(34) & 1.812(56) & 2.29(10)  \\
a09m130  & 3--20     & 0.4186(76) & 0.834(61) & 9.74(89)  & 17.2(1.0)  &         &            &           &           &           \\
\hline                                                                                                                
a06m310  & 4--30     & 0.3245(30) & 0.617(18) & 0.566(30) & 1.439(42)  & 4--30   & 0.3283(15) & 0.630(10) & 0.609(15) & 1.513(29) \\
a06m220  & 5--30     & 0.3166(66) & 0.644(54) & 13.0(1.5) & 38.5(5.4)  &         &            &           &           &           \\
\end{tabular}
\end{ruledtabular}
\caption{The nucleon ground and first excited state masses and the
  corresponding amplitudes obtained from a two-state fit to the
  nucleon two-point correlation function on each ensemble. The second
  set of estimates on the right are from an independent calculation
  performed to calculate the disconnected diagrams using the AMA with
  64 LP measurements (96 LP for {\it a06m310}).  All errors are
  estimated using the single-elimination Jackknife method using
  uncorrelated fits.}
\label{tab:res2pt}
\end{table*}

\subsection{Renormalization of Operators}
\label{sec:renorm}

The calculation of the renormalization constants $Z_\Gamma$ of the quark bilinear
operators in the RI-sMOM scheme~\cite{Martinelli:1994ty,Sturm:2009kb}
has been done on five ensembles: {\it a12m310, a13m220, a09m310,
  a09m220} and {\it a06m310}.  In order to translate the lattice
results to the continuum $\overline{\text{MS}}$ scheme at a fixed
scale, say $\mu = 2\GeV$, used by phenomenologists we follow the
procedure described in Ref.~\cite{Bhattacharya:2013ehc}.  To
summarize, the RI-sMOM estimate obtained at a given lattice
four-momentum $q^2$ is first converted to the $\overline{\text{MS}}$
scheme at the same scale (horizontal matching) using the one-loop
perturbative matching. This value is then run in the continuum in the
$\overline{\text{MS}}$ scheme to the fixed scale, $2\GeV$, using
the two-loop anomalous dimension.

Ideally, one would like to establish a window $\Lambda \ll q \ll c/a$
in the RI-sMOM scheme in which the $Z_\Gamma$ scales according to
perturbation theory.  Here $\Lambda$ is an infrared scale below which
nonperturbative effects are large and $c/a$ represents the cutoff
scale beyond which lattice discretization effects are large. The value
of $c$ is {\it a priori} unknown and the expectation is that it is
$O(1)$. Within this window, the scaling of $Z_T$ with $q^2$ gets
contributions from both the anomalous dimension of the operator and
the running of the strong coupling constant $\alpha_s$. If this
scaling is consistent with that predicted by perturbation theory, then
estimates within this window would converge to a constant value
independent of $q^2$ after conversion to $\overline{\text{MS}}$ scheme
and run to a fixed scale, $2\GeV$.  As discussed
in~\cite{Bhattacharya:2013ehc}, smearing the lattice to reduce the
ultraviolet noise in the measurements also reduces the upper cutoff
$c/a$ for the calculation of the renormalization constants, and 
{\it a priori}, we again do not know by how much smearing shrinks the
desired window or whether it totally eliminates it on the various
$0.06-0.12$~fm lattices we have analyzed.  Below we summarize the
tests performed and state the results.

\begin{figure}[tb]
  \subfigure{
    \includegraphics[width=0.95\linewidth]{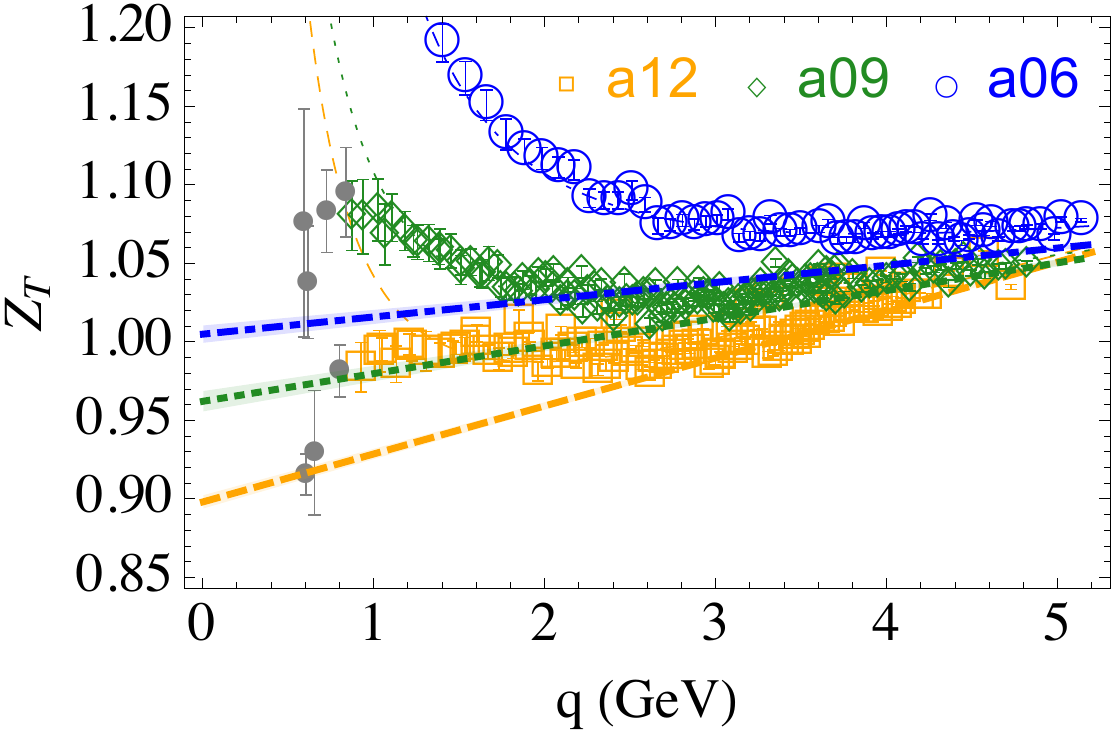}
  }
  \hspace{0.04\linewidth}
  \subfigure{
    \includegraphics[width=0.95\linewidth]{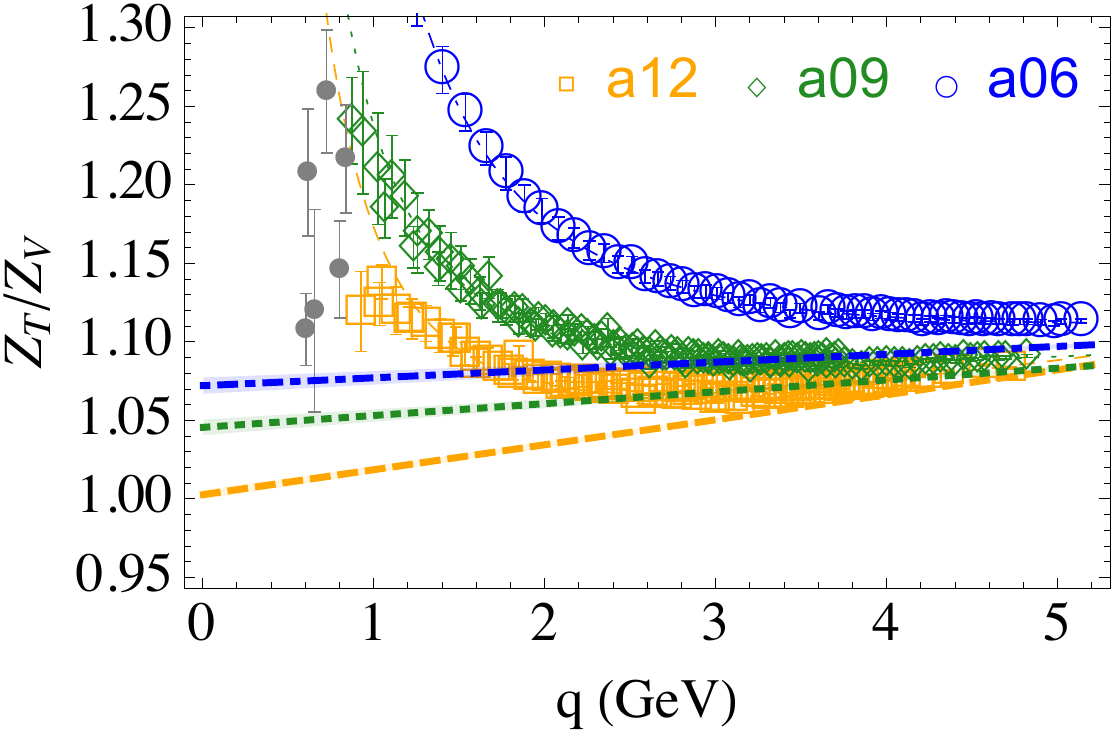}
  }
\caption{Data for $Z_T$ (upper) and $Z_T/Z_V$ (lower) after
  translation to the $\overline{\text{MS}}$ scheme at $2$~GeV as a
  function of the lattice momentum $q$ . The lattice calculation was
  done on five ensembles in the RI-sMOM scheme. The $a=0.12$~fm
  ($a=0.09$~fm) fit are to the combination of {\it a12m310} and 
  {\it a12m220} ({\it a12m310} and {\it a12m220}) data as there is no
  detectable dependence on the quark mass.  The $a=0.06$~fm fit is to
  the {\it a06m310} ensemble data. The data were fit using
  the ansatz $c/q^2 + Z + \alpha q $ and the dot-dashed, dotted and dashed lines show the
  extrapolation $Z + \alpha q $ for the $a=0.06$, $0.09$ and $0.12$~fm data.
  \label{fig:ZT_comp}}
\end{figure}

\begin{itemize}

\item
We first test the data for the $Z$'s in the RI-sMOM scheme to see if
they exhibit the desired perturbative behavior for HYP smeared
lattices by calculating the logarithmic derivative of $Z(q^2)$ and
comparing it to the anomalous dimension. The data show evidence of
such a window in the calculation of the vector, axial and tensor
renormalization constants, but not for the scalar. In this paper, we
only need $Z_T$ and $Z_V$, so we next describe how we obtained final
estimates for these and assigned a conservative error that covers the
various sources of systematic uncertainties.
\item
We find that the ratios of renormalization constants, $Z_\Gamma/Z_V$,
have less fluctuations and are flatter in $q^2$ as illustrated in
Fig.~\ref{fig:ZT_comp}. This improvement is presumably due to the
cancellation of some of the systematic uncertainties in the ratio,
including, for example, those due to the breaking of the continuum
Lorentz symmetry to the hypercubic rotation group on the lattice that
impacts the calculation of the $Z$-factors. On each ensemble, the
final renormalized charges can be constructed from these ratios as
$(Z_\Gamma /Z_V) \times (g_\Gamma/g_V^{u-d})$ using the identity $Z_V
g_V^{u-d} = 1$. Because of the better signal and resulting fits, we
use the estimates from the ratios method for our central values and
include the difference between these and estimates from the direct
calculation, $Z_\Gamma g_\Gamma$, as an estimate of the systematic
error.
\item
To take into account the remaining dependence on $q^2$ of the
estimates in $\overline{\text{MS}}$ scheme at $2\GeV$, we carry out
the two analysis strategies proposed in
Ref.~\cite{Bhattacharya:2013ehc}. In the first, we obtain the value
and error from the fit to the data using the ansatz $c/q^2 + Z +
\alpha q $. We find that these fits capture the data and the
extrapolations $Z + \alpha q $ are shown as dashed lines in
Fig.~\ref{fig:ZT_comp}.
\item
A slightly modified version of the second method is used: we now
choose the $q^2$ in the RI-sMOM scheme by the condition $q_i a -
\sin(q_i a) = 0.05$ based on bounding the discretization error and the
error in $Z$ is estimated from the spread in the data over a range in
$q^2$ about this point.  This choice corresponds to $q^2 = 5,\ 9$ and
$21$ GeV${}^2$ for the $a=0.12$, $0.09$ and $0.06$~fm ensembles,
respectively. The corresponding ranges for determining the error were
taken to be 4--6, 8--10 and 18--24 GeV${}^2$ over which the data show
a reasonably flat behavior as shown in Fig.~\ref{fig:ZT_comp}.
\item
For the final estimates we take the average of the two methods. The error 
is taken to be half the difference, and rounded up to be conservative. 
\item
On the ensembles at the two lattice spacings $a=0.12$ and $0.09$~fm,
we found no significant difference in the estimates of the
renormalization constants for the two different quark masses ($M_\pi =
310$ and $220$~MeV ensembles). A common fit captured both data sets,
as shown in Fig.~\ref{fig:ZT_comp}, and was used to extract our
``quark mass independent'' estimates.
\item
The entire calculation, matching to the $\overline{\text{MS}}$ scheme,
running to $2$~GeV and the final fits for the two strategies, was done
using 200 bootstrap samples because the number of configurations
analyzed in ensembles {\it a12m310} and {\it a12m220} ({\it a09m310}
and {\it a09m220}) are different.  
\end{itemize}

The final mass-independent renormalization constants at the three
lattice spacings needed to construct the renormalized charges in the
two ways: (i) $Z_\Gamma g_\Gamma$ and (ii) from the product of the
ratios $(Z_\Gamma /Z_V) \times (g_\Gamma/g_V^{u-d})$ with the identity
$Z_V g_V^{u-d} = 1$ are given in Table~\ref{tab:renorm}.  The errors
in the renormalization factors, $Z_T$ and $Z_T/Z_V$, are added in
quadratures to those in the extraction of the bare nucleon charges
$g_T^{\rm bare}$ and $g_T^\text{bare}/g_V^\text{bare}$, respectively,
to get the final estimates of the renormalized charges given in
Table~\ref{tab:res-renorm}.

\begin{table*}
\centering
\begin{ruledtabular}
\begin{tabular}{c|ccc|ccc|ccc}
ID   &  $Z_T$      & $Z_V$       & $Z_T/Z_V$  &  $Z_T$      & $Z_V$       & $Z_T/Z_V$   &  $Z_T$    & $Z_V$       & $Z_T/Z_V$ \\
\hline                                                                                                              
a12  & $0.898(4)$  & $0.890(4)$  & $1.003(3)$ & $0.995(10)$ & $0.918(12)$ & $1.073(2) $ & $0.95(5)$ & $0.90(2)$   & $1.04(4)$ \\
a09  & $0.962(6)$  & $0.911(9)$  & $1.045(5)$ & $1.026(6) $ & $0.938(6) $ & $1.089(2) $ & $0.99(4)$ & $0.925(15)$ & $1.07(3)$ \\
a06  & $1.005(6)$  & $0.931(4)$  & $1.072(5)$ & $1.071(5) $ & $0.961(5) $ & $1.1134(6)$ & $1.04(4)$ & $0.945(15)$ & $1.09(3)$ \\
\end{tabular}
\end{ruledtabular}
\caption{The mass independent renormalization constants $Z_T$, $Z_V$
  and the ratio $Z_T/Z_V$ in the $\overline{\text{MS}}$ scheme at
  $2\GeV$ at the three values of the lattice spacings used in our
  calculations. These estimates are obtained using the fit $1/q^2 + Z
  + \alpha q $ (left) and as an average over an interval in $q^2 $
  (middle) as described in the text. For the final estimates, shown in
  the last 3 columns, we take the average of the two methods and half the
  difference (rounded up) for the errors.}
\label{tab:renorm}
\end{table*}

\subsection{Statistical analysis of two-point and three-point functions}
\label{sec:statistics}

We carried out the following statistical analyses of the data on each
ensemble to look for anomalies. We divided the data for a given
ensemble into bins of about 1000 measurements (by source points and by
configuration generation order) to test whether the ensembles consist
of enough independent configurations. For bin sizes $ > 5$
configurations, the errors in the mean decreased as $\sqrt N$, $i.e.$,
consistent with our analysis of the autocorrelation coefficient of
about $5$ configurations (about 25 molecular dynamics steps). Also,
the error computed with data averaged over $S$ source points on each
configuration is smaller by $\sqrt S$ compared to the error in the data
from any one of the source position.

Estimates from bins of about 1000 measurements, however, fluctuated by
up to $3\sigma$ in some cases. This variation is much larger than
expected based on the bin sizes. To determine whether the data in the
various bins satisfy the condition of being drawn from the same
distribution, we performed the Kolmogorov–-Smirnov (K-S) test on
quantities that have reasonable estimates configuration by
configuration, for example, the isovector vector charge $g_V^{u-d}$
and the value of the two-point function at a given time separation. The
K-S test showed acceptable probability of the various bins being drawn
from the same distribution. Histograms of the data showed no long
tails in the distribution but exhibit variations in the sample
distribution that becomes increasingly Gaussian as the bin size was
increased to the full sample size. 

We find these $2-3\sigma$ fluctuations both when the data are binned 
by the source position and when the configurations are divided into
two halves according to the molecular dynamics generation order.  Such
fluctuations are apparent in the {\it a06m310} and the {\it a06m220}
ensemble data. Comparing the data for different charges (axial, scalar
and tensor), we found that the effect is least significant (less than
$1\sigma$) for the tensor charge and worst for the vector charge
$g_V$; it is, presumably, most evident in $g_V$ because it has the
smallest statistical errors. We offer two possible explanations.  One,
the large variation observed in the bin mean indicates that the
ensembles of $O(1000)$ configurations (spanning a total of 5000-6000
molecular dynamics evolution steps in the generation of thermalized
HISQ lattices we have used) have not covered enough phase space and
bin errors are consequently underestimated. The other explanation is
that, since we used the same four or eight source positions on all
configurations in an ensemble, the data for fixed source position is
more correlated. Our ongoing tests confirm that using random but
well-separated source positions on each configuration is a better
strategy. Finally, based on the convergence of the bin distributions
to a gaussian on increasing the bin size to the full sample and
the lack of evidence of long tails, makes us confident that the final
error estimates are reliable.

Our overall conclusion about statistics is that while $O(10,000)$
measurements on these ensembles of $O(1000)$ configurations are
sufficient for extracting the tensor charge with few percent
uncertainty, one will need a factor of ten or more in statistics for
obtaining the scalar charge with similar accuracy.  This goal is currently 
being pursued using the AMA method discussed in Section~\ref{sec:disc}. 

Lastly, we performed both correlated and uncorrelated fits to the nucleon
two-point function data. In all cases in which the correlated fits
were stable under changes in the fit ranges and had reasonable $\chi^2$,
the two fits gave overlapping estimates.  Since correlated fits did
not work in all cases, all statistical errors in the two- and
three-point correlation functions were, thereafter, calculated using a
single elimination jackknife method with uncorrelated fits performed
on each jackknife sample.

\section{Contribution of the Connected Diagram}
\label{sec:connected}

Estimates of the bare and renormalized charges on the nine ensembles
at different lattice spacings, light quark masses and lattice
volumes are given in Tables~\ref{tab:res} and~\ref{tab:res-renorm}.

\begin{figure*}[tb]
  \subfigure{
    \includegraphics[width=0.95\linewidth]{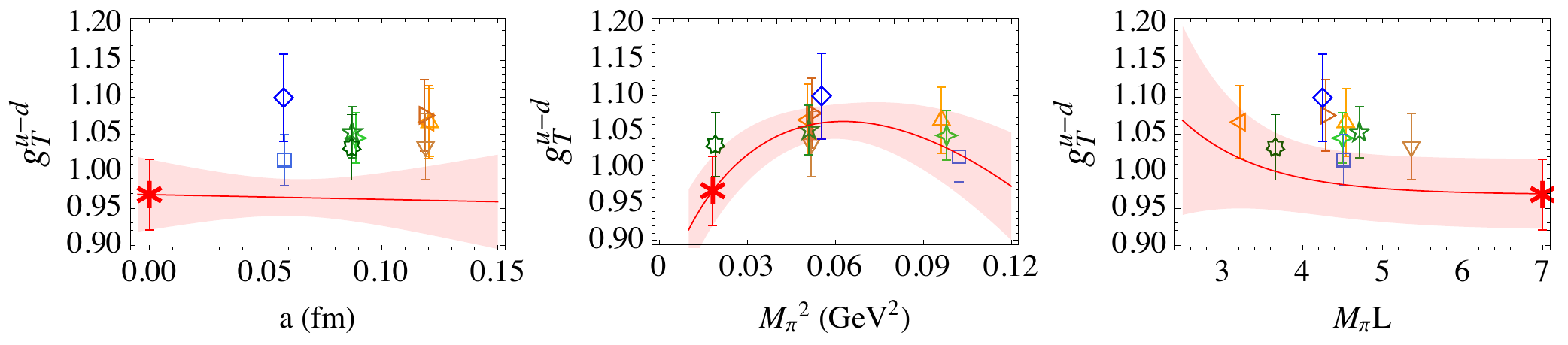}
  }
  \hspace{0.04\linewidth}
  \subfigure{
    \includegraphics[width=0.95\linewidth]{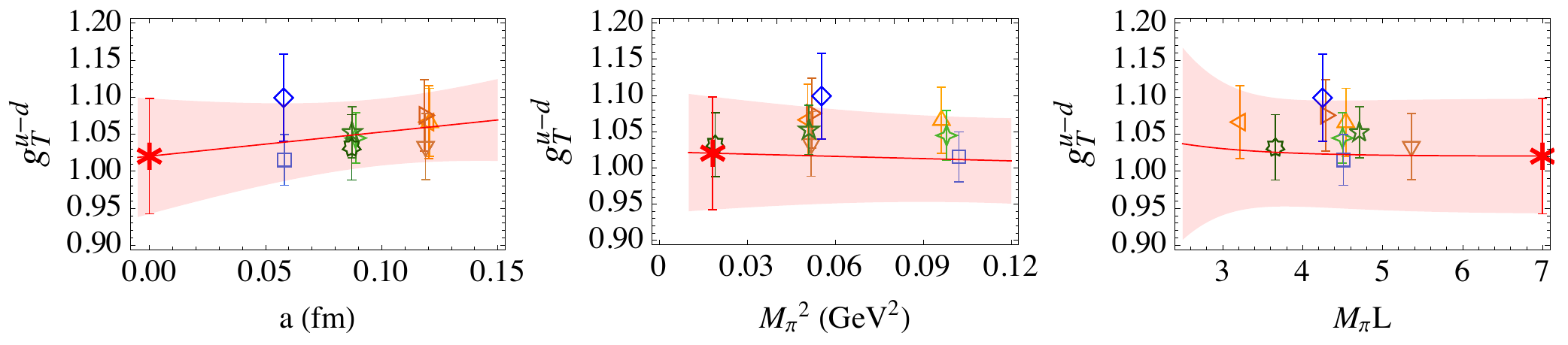}
  }
\caption{Comparision of the simultaneous fits versus $a$, $M_\pi^2$
  and $M_\pi L$ to the isovector charge, $g_T^{u-d}$, data using
  Eq.~\protect\eqref{eq:chiralfit} (top) with the simpler version
  without the chiral logarithms given in Eq.~\protect\eqref{eq:extrap}
  (bottom).  The data symbols are defined in Table~\ref{tab:ens}. The
  fit is given by the red line and the physical value after
  extrapolation to the continuum limit ($a\rightarrow 0$), physical
  pion mass ($M_\pi \rightarrow M_{\pi^0}^{{\rm phys}}$) and infinite
  volume ($L \rightarrow \infty$) is marked by a red star. The error
  band is shown as a function of each variable holding the other two
  at their physical value.  The data are shown projected on to each of
  the three planes.
  \label{fig:chiral_gT_compare}}
\end{figure*}

To extrapolate these estimates to the physical point, $i.e.$, the
continuum limit ($a\rightarrow 0$), the physical pion mass ($M_{\pi^0}
= 135$~MeV) and the infinite volume limit ($L \rightarrow \infty$), we
explored the four parameter ansatz
\begin{align}
  g_T^{i} =  \ c_1^{i}  \left[ 1 +   \frac{M_\pi^2}{(4 \pi  F_\pi)^2}   \ f^{i}  \left( \frac{M_\pi}{\mu} \right) \right] \nonumber \\ 
       +   c_2^{i}  \, a +  c_3^{i}  (\mu)   \, M_\pi^2  + c_4^{i} \, e^{-M_\pi L} ~.
\label{eq:chiralfit}
\end{align}
where we have included the leading chiral logarithms~\cite{deVries:2010ah}. 
The loop functions $f^{i}(\mu/M_\pi)$ for the two isospin channels are: 
\begin{align}
  f^{u+d} &=&  \frac{3}{4} \left[   \left( 2 + 4 g_A^2 \right)   \log \frac{\mu^2}{M_\pi^2}    + 2 + g_A^2   \right]  
\nonumber \\ 
&=& 2.72 + 6.38 \log \frac{\mu^2}{M_\pi^2}  ~, 
\\
  f^{u-d} &=&  \frac{1}{4} \left[   \left( 2 + 8  g_A^2 \right)   \log \frac{\mu^2}{M_\pi^2}    + 2 + 3 g_A^2   \right] 
\nonumber \\ 
&=& 1.72 + 3.75 \log \frac{\mu^2}{M_\pi^2}  ~,
\label{eq:chiral_u-d}
\end{align}
where we use $\mu = M_\rho = 770$~MeV for the renormalization scale
and $g_A = 1.276$.  The extrapolation ansatz is taken to be linear in
$a$ because the discretization errors in the clover-on-HISQ formalism
with unimproved operators start at $\mathcal{O}(a)$. Similarly, we
have kept only the leading finite volume correction term,
$e^{-M_\pi L}$.  In Fig.~\ref{fig:chiral_gT_compare}, we compare the fit
obtained using Eq.\eqref{eq:chiralfit} with that using the simpler
isospin independent four parameter ansatz without the chiral logarithm:
\begin{align}
  g_T (a,M_\pi,L) = c_1 + c_2a + c_3 M_\pi^2 + c_4 e^{-M_\pi L} \,.
\label{eq:extrap}
\end{align}
Both fits have reasonable $\chi^2/$dof and the estimates at the
physical point are consistent. The fit including the chiral logarithm would 
naively indicate that $g_T$ should decrease in value with increasing
$M_\pi^2$ for $M_\pi > 300$~MeV. Such a behavior is not seen in the
global data shown in~Fig.~\ref{fig:GlobalData}. We conclude that the large
curvature due to the chiral logarithm seen in
Fig.~\ref{fig:chiral_gT_compare} is most likely due to the 
number and accuracy of the data and of keeping just the leading chiral
correction.  Also, the error estimate from the fit using the simpler
ansatz given in Eq.\eqref{eq:extrap} is more conservative and covers
the full range of both fits.  We, therefore, use Eq.~\eqref{eq:extrap}
for all further analysis in this paper.

\begin{figure*}[tb]
  \subfigure{
    \includegraphics[width=0.95\linewidth]{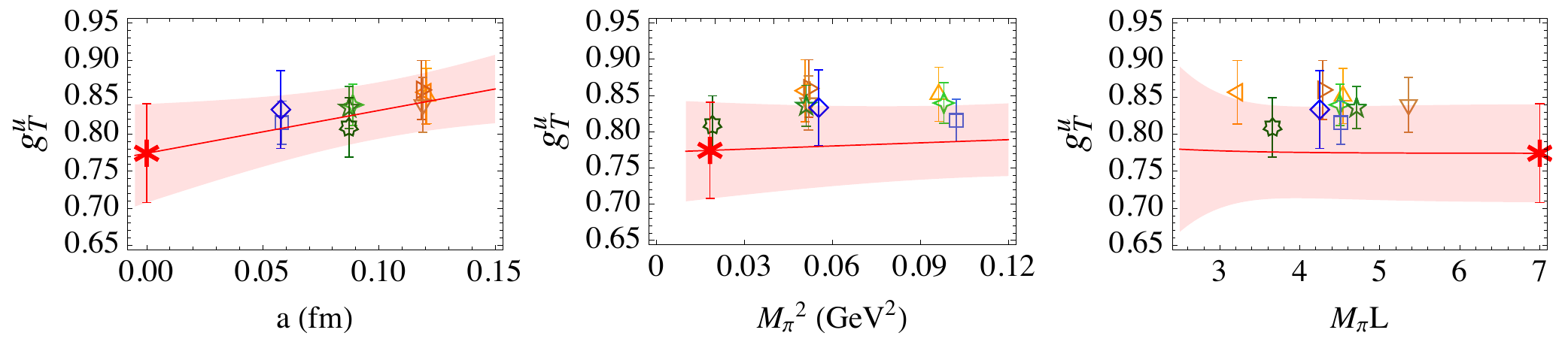}
  }
  \hspace{0.04\linewidth}
  \subfigure{
    \includegraphics[width=0.95\linewidth]{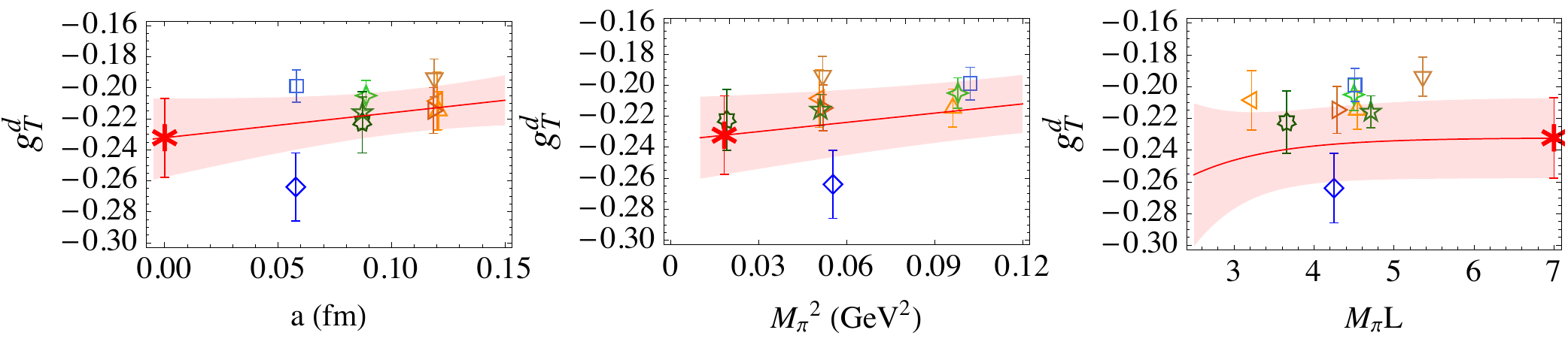}
  }
\caption{Simultaneous extrapolation to the physical point
  ($a\rightarrow 0$, $M_\pi \rightarrow M_{\pi^0}^{{\rm phys}}$, and
  $L \rightarrow \infty$) using Eq.~\protect\eqref{eq:extrap}, of the
  connected contributions to the flavor diagonal nucleon (proton) tensor
  charges, $g_T^u$ (upper) and $g_T^d$ (lower), renormalized in the
  $\overline{{\rm MS}}$ scheme at $2\GeV$.  The physical values given
  by the fit are marked by a red star.  The rest is the same as in
  Fig.~\protect\ref{fig:chiral_gT_compare}.
  \label{fig:con_extrap}}
\end{figure*}

\begin{figure*}[tb]
  \subfigure{
    \includegraphics[width=0.98\linewidth]{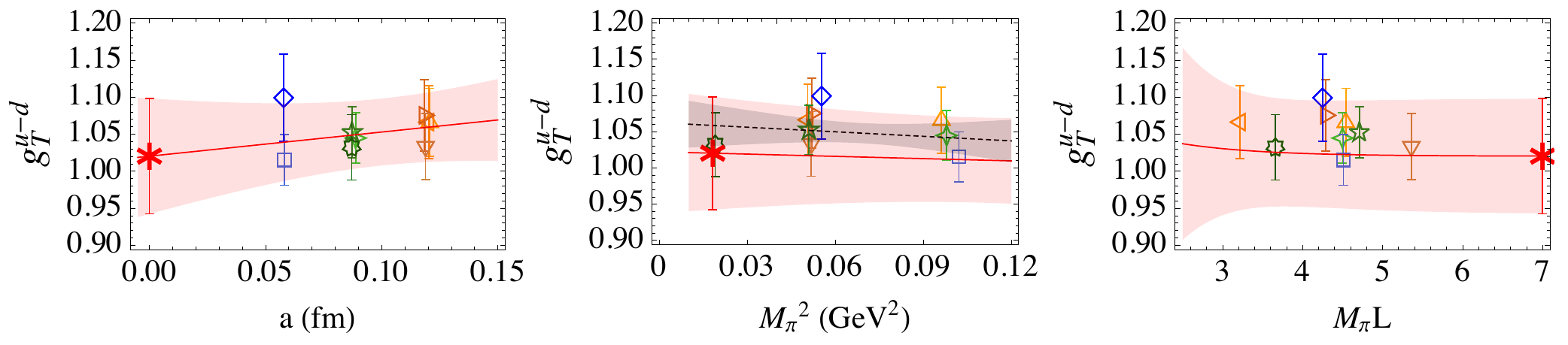}
  }
  \hspace{0.04\linewidth}
  \subfigure{
    \includegraphics[width=0.98\linewidth]{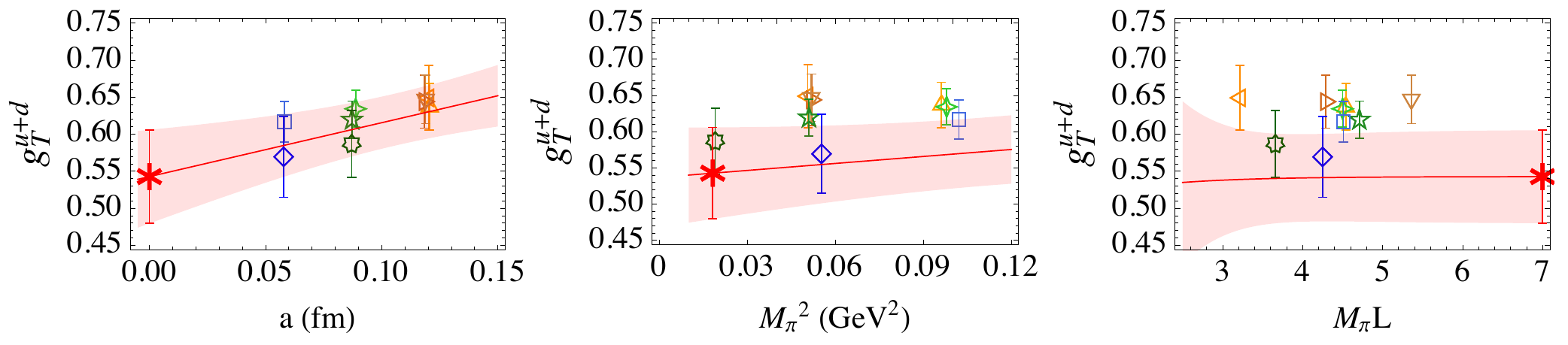}
  }
\caption{Simultaneous extrapolation to the physical point
  ($a\rightarrow 0$, $M_\pi \rightarrow M_{\pi^0}^{{\rm phys}}$, and
  $L \rightarrow \infty$) using Eq.~\protect\eqref{eq:extrap} of the
  connected contributions to the isovector $g_T^{u-d}$ (upper) and
  isoscalar $g_T^{u+d}$ (lower) nucleon (proton) tensor charges renormalized in
  the $\overline{{\rm MS}}$ scheme at $2\GeV$. The overlay in the
  middle upper figure, with the dashed line within the thin grey band, is
  the fit to the data versus $M_\pi^2$ assuming no dependence on the
  other two variables.  Rest is the same as in
  Fig.~\protect\ref{fig:chiral_gT_compare}.
  \label{fig:conUD_extrap}}
\end{figure*}

The results of the fits using Eq.\eqref{eq:extrap} and the
extrapolated value are shown in Fig.~\ref{fig:con_extrap} separately
for operator insertion on the $u$ and $d$ quarks in the nucleon.  We
find that the $g^u_T$ contribution is larger and essentially flat in
all three variables (lattice spacing, pion mass and volume), while the
$g^d_T$ connected contribution is much smaller and shows a slightly
larger relative spread. The spread in $g_T^d$ on the $a=0.06$~fm
lattices is an example of the unexpectedly large statistical
fluctuations we mentioned in section~\ref{sec:statistics} that will
require higher statistics to resolve.  The final renormalized
extrapolated values for the proton charges are
\begin{align}
g_T^{u}(con) & = 0.774(65) \,, \nonumber \\
g_T^{d}(con) & = -0.233(25) \,.
\label{eq:conn_extrap}
\end{align}
The $\chi^2/\text{dof}$ is 0.1 and 1.6 for $g_T^u$ and $g_T^d$,
respectively, with $\text{dof}=5$.  In performing the fits, we assume
that the error in each data point has a Gaussian distribution even
though the quoted $1\sigma$ error is a combination of the statistical
error and the systematic error coming from the calculation of the
renormalization factor $Z_T$.
The fits to the
isovector $g_T^u-g_T^d$ and the connected part of the isoscalar $g_T^u+g_T^d$ data 
using Eq.~\eqref{eq:extrap}, are shown in Fig.~\ref{fig:conUD_extrap}.
Our final estimates are
\begin{align}
g_T^{u-d}(con) & =  1.020(76) \,, \nonumber \\
g_T^{u+d}(con) & =  0.541(62) \,.
\label{eq:conn_extrap2}
\end{align}
with a $\chi^2/dof = 0.4$ and $0.2$ respectively.

\begin{table*}
\centering
\begin{ruledtabular}
\begin{tabular}{c|cccc|cc|c}
ID       & $g_T^{{\rm con}, u}$ & $g_T^{{\rm con}, d}$ & $g_T^{{\rm con}, u-d}$ & $g_T^{{\rm con}, u+d}$ & $g_T^{{\rm disc}, l}$ & $g_T^{{\rm disc}, s}$ & $g_V^{{\rm con}, u-d}$ \\ 
\hline
a12m310  & 0.875(18) & $-$0.2208(93) & 1.096(21) & 0.655(20) & $-$0.0124(23) & $-$0.0041(20) & 1.069(9) \\
a12m220S & 0.873(26) & $-$0.212(17)  & 1.086(26) & 0.661(36) & ---           & ---           & 1.059(12) \\
a12m220  & 0.888(22) & $-$0.222(12)  & 1.111(24) & 0.665(26) & $-$0.0038(41) & $-$0.0010(28) & 1.074(11)  \\
a12m220L & 0.859(18) & $-$0.198(10)  & 1.058(19) & 0.662(21) & ---           & ---           & 1.065(7) \\
\hline
a09m310  & 0.829(16) & $-$0.2025(80) & 1.031(19) & 0.626(18) & $-$0.0050(21) & $-$0.0005(21) & 1.056(8) \\
a09m220  & 0.820(16) & $-$0.2120(79) & 1.033(17) & 0.608(18) & ---           & $-$0.0021(53) & 1.050(9) \\
a09m130  & 0.779(33) & $-$0.214(18)  & 0.993(33) & 0.565(42) & ---           & ---           & 1.029(16)  \\
\hline
a06m310  & 0.778(18) & $-$0.1898(86) & 0.969(20) & 0.588(21) & $-$.0035(62)  & $-$0.0005(52) & 1.040(8) \\
a06m220  & 0.759(43) & $-$0.241(19)  & 1.002(46) & 0.519(48) & ---           & ---           & 0.993(18) \\
\end{tabular}
\end{ruledtabular}
\caption{The bare connected ($g_T^{\rm con}$) and disconnected
  ($g_T^{{\rm disc}}$) contributions to the tensor charges of the
  proton on the nine ensembles.  Dashes indicate that those ensembles
  have not been simulated. The isovector vector charge $g_V^{u-d}$ is
  used to construct ratios for noise reduction as described in the
  text.}
\label{tab:res}
\end{table*}

\begin{table*}
\centering
\begin{ruledtabular}
\begin{tabular}{c|cccc|cc}
ID       & $g_T^{{\rm con}, u}$ & $g_T^{{\rm con}, d}$ & $g_T^{{\rm con}, u-d}$ & $g_T^{{\rm con}, u+d}$ & $g_T^{{\rm disc}, l}$ & $g_T^{{\rm disc}, s}$ \\
\hline
a12m310  & 0.852(37) & $-$0.215(12)  & 1.066(46) & 0.637(31) & $-$0.0121(23) & $-$0.0040(19) \\
a12m220S & 0.857(43) & $-$0.209(19)  & 1.066(50) & 0.649(44) & ---           & ---           \\
a12m220  & 0.860(40) & $-$0.215(15)  & 1.075(48) & 0.644(36) & $-$0.0037(40) & $-$0.0010(27) \\
a12m220L & 0.840(37) & $-$0.194(12)  & 1.033(45) & 0.647(33) & ---           & ---           \\ 
\hline
a09m310  & 0.840(28) & $-$0.2051(98) & 1.045(34) & 0.634(25) & $-$0.0050(22) & $-$0.0005(21) \\
a09m220  & 0.836(28) & $-$0.216(10)  & 1.053(34) & 0.619(25) & ---           & $-$0.0021(54) \\
a09m130  & 0.809(40) & $-$0.222(20)  & 1.032(44) & 0.587(45) & ---           & ---           \\
\hline
a06m310  & 0.815(29) & $-$0.199(10)  & 1.015(34) & 0.617(27) & $-$0.0037(65) & $-$0.0005(55) \\
a06m220  & 0.833(52) & $-$0.264(22)  & 1.099(59) & 0.569(55) & ---           & ---           \\
\end{tabular}
\end{ruledtabular}
\caption{The renormalized connected ($g_T$) and disconnected
  ($g_T^{{\rm disc}}$) contributions to the tensor charges of the
  proton on the nine ensembles. The errors are obtained by adding in
  quadratures the statistical errors given in Table~\ref{tab:res} in
  the bare charges to the errors in the renormalization constants
  given in Table~\ref{tab:renorm}.}
\label{tab:res-renorm}
\end{table*}

%

\section{Contribution of the Disconnected Diagram}
\label{sec:disc}

In Section~\ref{sec:methodology} we showed that to estimate the
disconnected contribution, we need to calculate two quantities at
zero-momentum---the nucleon two-point function and the contraction of
the bilinear fermion operator into a quark loop---and measure their
correlation.  These two calculations are described below.

\subsection{Two-point Function}
\label{sec:2ptD}

The high statistics calculation of the two-point function with smeared
sources was redone using the all-mode-averaging (AMA)
technique~\cite{Blum:2012uh} because quark propagators from the
earlier connected three-point function study were too expensive to
store.
To implement AMA, we again choose four different source time slices
separated by $L_T/4$ on each configuration. On each of these
time slices we calculate the two-point correlator by placing
$N_{\rm LP}=15$ low precision (LP) sources, for a total of $4\times
15=60$ sources per configuration. This estimate is {\it a priori}
biased due to the LP calculation. In addition, on each of these four
time slices we place one high precision (HP) source, $i.e.$,
$N_{\rm HP}=4$ such sources per configuration, from which we
calculate a LP and a HP correlator.  These four HP and LP correlators
are used to correct the bias in the 60 LP estimates, $i.e.$, on each
configuration, the two-point function is given by
\begin{align}
 C^\text{2pt, imp}&(t, t_0) 
 = \frac{1}{N_\text{LP}} \sum_{i=1}^{N_\text{LP}} 
    C_\text{LP}^\text{2pt}(t; t_0, \mathbf{x}_i^\text{LP}) \nonumber \\
  +& \frac{1}{N_\text{HP}} \sum_{i=1}^{N_\text{HP}} \left[
    C_\text{HP}^\text{2pt}(t; t_0, \mathbf{x}_i^\text{HP})
    - C_\text{LP}^\text{2pt}(t; t_0, \mathbf{x}_i^\text{HP})
    \right] \,,
  \label{eq:2pt_est}
\end{align}
where $C_\text{LP}^\text{2pt}$ and $C_\text{HP}^\text{2pt}$ are the
two-point correlation function calculated in LP and HP, respectively,
and $\mathbf{x}_i^\text{LP}$ and $\mathbf{x}_i^\text{HP}$ are the two
kinds of source positions. 

The basic idea of AMA is that, in the low-precision evaluation, the LP
average (first term in Eq.~\eqref{eq:2pt_est}) is biased, and this
bias depends predominately on low modes of the Dirac operator that are
independent of the source position and can be corrected by the second
term. Thus, we get an unbiased estimate from 60 LP source points for
the computational cost of ($60+4$) LP and 4 HP calculations. In our
current implementation, 15 LP measurements cost the same as one HP
when using the multigrid algorithm for inverting the Dirac
matrix~\cite{Babich:2010qb}. (On the {\it a06m310} ensemble we used
(92+4) LP and 4 HP sources and the errors decreased by a factor of
$\sim 1.2$ compared to (60+4) LP sources.) Comparing the errors in the
estimates for masses given in Table~\ref{tab:res2pt}, we find that the
AMA errors are a factor of $2-4$ times smaller than those from the
connected study (all HP measurements). Since this improvement is based
on comparing 120 LP (we effectively doubled the LP statistics by
analyzing both the forward and backward propagation of the nucleon)
versus 8 HP measurements, we conclude that the correlations between
the LP measurements are small.

If we assume that the variance of both the LP and HP measurements is
the same and given by $\sigma$ and the correlation between the HP and
LP measurements from the $N_{\rm HP}$ points, ${\cal C} =
\sigma^2_{NP,LP}/\sigma^2$, is small, then the statistical error in 
Eq.~\eqref{eq:2pt_est} is given by \cite{Blum:2012uh}
\begin{align}
\label{eq:err_imp}
 \sigma^\text{imp} \approx
 \sigma \sqrt{\frac{1}{N_\text{LP}} + \frac{2}{N_\text{HP}}(1-{\cal C})}\,.
\end{align}
The second term in the square root becomes smaller as the LP estimate 
approaches the HP estimate and the correlation factor ${\cal C} \rightarrow 1$.
By controlling $N_\text{LP}$, $N_\text{HP}$ and ${\cal C} $, we can
minimize the total error for a fixed computational cost.

To speed up the AMA method we exploit the fact that the same Dirac
matrix is inverted multiple times\footnote{The number of inversions of
  the Dirac matrix per configuration are $12\times(N_\text{LP}+
  N_\text{HP})$} on each configuration. It is, therefore, efficient to
precondition the matrix by deflating the low-eigenmodes.
We implement such improvement using the multigrid solver~\cite{Babich:2010qb, Osborn:2010mb}
which has deflation built in.
To obtain the LP estimate of the two-point function, we truncate the
multigrid solver using a low-accuracy stopping criterion: the ratio
($r_{\rm LP} \equiv |{\rm residue}|_{\rm LP}/|{\rm source}|$) is
chosen to be $10^{-3}$ for all the ensembles.  Our final analysis of
the masses, amplitudes and matrix elements, however, shows that this
stopping criteria was overly conservative as the bias correction term
is negligible compared to the statistical errors.

\subsection{Disconnected Quark Loop}
\label{sec:DQLC}

For the evaluation of the quark loop term 
$\sum_\mathbf{x} \Tr [M^{-1}(\tau, \mathbf{x}; \tau, \mathbf{x}) \Gamma]$,
we adopt the stochastic method accelerated with a combination of the truncated solver method (TSM)
\cite{Collins:2007mh, Bali:2009hu}, the hopping parameter expansion (HPE) 
\cite{Thron:1997iy, Michael:1999rs} and the dilution technique 
\cite{Bernardson:1994at, Viehoff:1997wi, Foley:2005ac}.
To obtain a stochastic estimate of the quark loops, consider a set of
random complex noise vectors $\vert \eta_i \rangle$ for
$i=1,2,3,\cdots,N$, having color, spin and spacetime components with
the following properties:
\begin{align}
 \frac{1}{N} \sum_{i=1}^N \vert \eta_i \rangle 
  &= \mathcal{O}\left(\frac{1}{\sqrt{N}}\right) \,, \\
 \frac{1}{N} \sum_{i=1}^N \vert \eta_i \rangle \langle \eta_i \vert 
  &= \mathbb{1} + \mathcal{O}\left(\frac{1}{\sqrt{N}}\right) \,.
\end{align}
We choose complex Gaussian noise vectors, \emph{i.e.}, we fill all the
spin, color and spacetime components of the vector with $(r_r +
ir_i)/\sqrt{2}$, where $r_r$ and $r_i$ are Gaussian random numbers,
because they give marginally smaller statistical error than
$\mathbb{Z}_N$ random noise when combined with the HPE.

These random vectors are used as sources for the inversion of the Dirac matrix. 
Then, from the solutions $\vert s_i \rangle$ of the Dirac equation,
\begin{align}
\label{eq:dirac_eq}
 M \vert s_i \rangle = \vert \eta_i \rangle \,,
\end{align}
the inverse of the Dirac matrix is given by 
\begin{align}
 M^{-1} 
  &= \frac{1}{N} \sum_{i=1}^N \vert s_i \rangle \langle \eta_i \vert
    + M^{-1} \left(\mathbb{1} - \frac{1}{N} \sum_{i=1}^N 
      \vert \eta_i \rangle \langle \eta_i \vert\right) 
  \label{eq:stoch_prop_a}\\
  &= \frac{1}{N} \sum_{i=1}^N \vert s_i \rangle \langle \eta_i \vert 
    + \mathcal{O}\left( \frac{1}{\sqrt{N}} \right)\,.
  \label{eq:stoch_prop}
\end{align}
The stochastic estimate of the zero-momentum insertion of the operator 
contracted into a quark loop is then given by 
\begin{align}
\label{eq:stoch_eval}
 \sum_\mathbf{x} \Tr [M^{-1}(\tau, \mathbf{x}; \tau, \mathbf{x}) \Gamma]
 \approx \frac{1}{N} \sum_{i=1}^N \langle \eta_i \vert_\tau \Gamma \vert s_i \rangle_\tau 
 \,,
\end{align}
where $\vert \eta \rangle_\tau$ is a vector whose $t=\tau$ components are
filled with random numbers, and the entries on other time slices are
set to zero.
%

In the estimation of the inverse of the Dirac matrix by 
using random sources, Eq.~\eqref{eq:stoch_prop}, one can use the mixed-precision technique called the 
truncated solver method (TSM) \cite{Collins:2007mh, Bali:2009hu}.
The idea of the TSM is the same as the AMA used in the evaluation of
the two-point function.
Consider two kinds of solution vectors of 
Eq.~\eqref{eq:dirac_eq} for a given random source $\vert \eta_i \rangle$ with different precision:
$\vert s_i \rangle_\text{LP}$ and $\vert s_i \rangle_\text{HP}$, where 
$\vert s_i \rangle_\text{LP}$ is the low precision computationally cheap estimate of the 
solution, while $\vert s_i \rangle_\text{HP}$ is the high precision solution.
The low precision estimate, $\vert s_i \rangle_\text{LP}$, was
obtained by truncating the multigrid inverter at $r_\text{LP} =
5\times 10^{-3}$.
The bias observed with this choice of $r_\text{LP}$ will be discussed later in this section.

By using the LP and HP solutions, the unbiased estimator of $M^{-1}$ is again given by 
\begin{align}
M_E^{-1} 
  &= \frac{1}{N_\text{LP}} \sum_{i=1}^{N_\text{LP}} 
      \vert s_i \rangle_\text{LP} \langle \eta_i \vert \nonumber \\
   &+ \frac{1}{N_\text{HP}} \sum_{i=N_\text{LP}+1}^{N_\text{LP} +N_\text{HP}} 
      \left( \vert s_i \rangle_\text{HP} 
        - \vert s_i \rangle_\text{LP}\right) \langle \eta_i \vert \,.
\label{eq:stoch_m_inv}
\end{align}
The first term in the right-hand-side (r.h.s) is the LP estimate of $M^{-1}$
while the second term in the r.h.s corrects the bias. 
As described in the case of the two-point function estimation, the total statistical
error of $M_E^{-1}$ scales as Eq.~\eqref{eq:err_imp}.
In other words, there are again two sources of statistical error in $M_E^{-1}$:
one is the LP estimate that scales as ${\sqrt{1/N_\text{LP}}}$, and the other is
the correction term that scales as ${\sqrt{1/N_\text{HP}}}$.  The size of the 
statistical error in the correction term is determined by the correlation 
between $\vert s_i \rangle_\text{HP}$ and $\vert s_i \rangle_\text{LP}$.

In the TSM, there are three parameters we can tune to minimize the statistical error 
for a given computation cost: $N_\text{LP}$, 
$N_\text{LP}/N_\text{HP}$ and the LP stopping criteria $r_\text{LP}$.
Note that once $N_\text{LP}/N_\text{HP}$ and $r_\text{LP}$ are determined, the total 
error scales as ${\sqrt{1/N_\text{LP}}}$.
Hence $N_\text{LP}$ determines the size of the error, and $N_\text{LP}/N_\text{HP}$ 
and $r_\text{LP}$ determine the efficiency of the estimator in terms of the 
computational cost.
To maximize the efficiency, we tune the $N_\text{LP}/N_\text{HP}$ and
$r_\text{LP}$ so that the size of the error from the correction term is
much smaller as it minimizes the computation time.
In this study, we use $N_\text{LP}/N_\text{HP}=30$ or 50 (See
Table.~\ref{tab:params_disc}) and $r_\text{LP} \sim 5\times 10^{-3}$.
With this accuracy, we find that the bias correction term is $\sim
10\%$ of the final estimate of $g_T^l(disc)$ and about half of the
statistical error.

We improve the TSM by using the hopping parameter expansion
(HPE)~\cite{Thron:1997iy, Michael:1999rs} as a preconditioner to
reduce the statistical noise. In the HPE one writes the clover Dirac matrix 
as 
\begin{align}
 M = \frac{1}{2\kappa}\left(\mathbb{1} - \kappa D \right)
 \,,
\end{align}
where $\kappa$ is the hopping parameter. The inverse can then be written as 
\begin{align}
 \frac{1}{2\kappa} M^{-1} 
  = \mathbb{1} + \sum_{i=1}^{n-1} (\kappa D)^i + (\kappa D)^n \frac{1}{2\kappa} M^{-1}
\,.
\end{align}
By taking $n=2$, the disconnected quark loop is given by 
\begin{align}
 \Tr\left[M^{-1} \Gamma \right] 
 = \Tr\left[\left(2\kappa \mathbb{1} + 2\kappa^2 D 
    + \kappa^2 D^2 M^{-1}\right) \Gamma \right]
\,.
\end{align}
Here, the first two terms of the r.h.s do not contribute to the nucleon
tensor charge because $\Tr \Gamma = \Tr (\Gamma \gamma_\mu) = 0$. 
%
As a result, the only nontrivial term that we need to calculate is
$\Tr\left[\kappa^2 D^2 M^{-1} \Gamma \right]$.
Because the two leading terms, which would otherwise contribute only
to the noise, are removed from the stochastic estimation, HPE works as
an error reduction technique.
Tests using the {\it a12m310} ensemble show that the statistical error
of the disconnected contribution to the tensor charge is reduced by a
factor of about 2.5 with HPE.
%

As shown in Eq.~\eqref{eq:stoch_prop_a}, the noise in the stochastic estimation
for $M^{-1}$ is proportional to $M^{-1}$, whose magnitude decreases 
exponentially as the spacetime distance between source and sink increases.
Hence it is possible to reduce the statistical noise by placing noise
sources only on part of the whole time slice, choosing maximally
separated points, and fill the other points on the time slice with zero.
This procedure divides the time slice into $m$ subspaces, and the
answer for the full time slice is obtained by combining results of
the $m$ subspaces.
The computational cost increases by a factor of $m$ because Dirac inversions
are needed for each noise source vector defining a subspace.
Hence this technique is useful when the reduction in noise wins over
the increase in computational cost.
This is called the time dilution method~\cite{Bernardson:1994at, Viehoff:1997wi, 
Foley:2005ac}.
Unfortunately, we find that the increase in computational cost is
equal to or larger than the gain from the reduction of statistical
noise for the nucleon charges.
Hence we place random sources on all points of a time slice and for 
each time slice that we want to evaluate the operator on.

There is one more symmetry that can be used for noise reduction:
$\gamma_5$-hermiticity of clover Dirac operator, $M^\dag = \gamma_5 M \gamma_5$.
Because of this symmetry, the quark loop for tensor channel should be pure
imaginary, and the nucleon two-point function is real. 
Hence we set the real part of the quark loop to zero when constructing
the correlation function and averaging over the configurations in
Eq.~\eqref{eq:3pt_2pt_ratio_disc}.

To increase the statistics, we average over the three possible
combinations of $\gamma_i\gamma_j$ and forward/backward propagators.
The final values of $t_{\rm sep}$ investigated, the displacement $\tau$ with respect to the source time slice of the
two-point correlator on which the operator was inserted, the
statistics and the number of random noise sources used on each
configuration are given in Table~\ref{tab:params_disc}.

\subsection{Results for the disconnected contributions}
\label{sec:resultsD}

The calculation of the disconnected diagram is computationally
expensive so it has been done on only four ensembles: 
{\it a12m310, a12m220, a09m310} and {\it a06m310}.
These four ensembles provide an understanding of the discretization
errors and of the behavior as a function of the quark mass. To get the
full contribution of the quark EDM to the nucleon EDM, we also need to
evaluate the disconnected diagram with a strange quark loop. Since the
calculation with the strange quark are computationally cheaper, we
have also analyzed a fifth ensemble, {\it a09m220}, for that estimate.

\begin{figure*}[tb]
  \begin{flushleft}
   \subfigure{
    \includegraphics[width=0.33\linewidth]{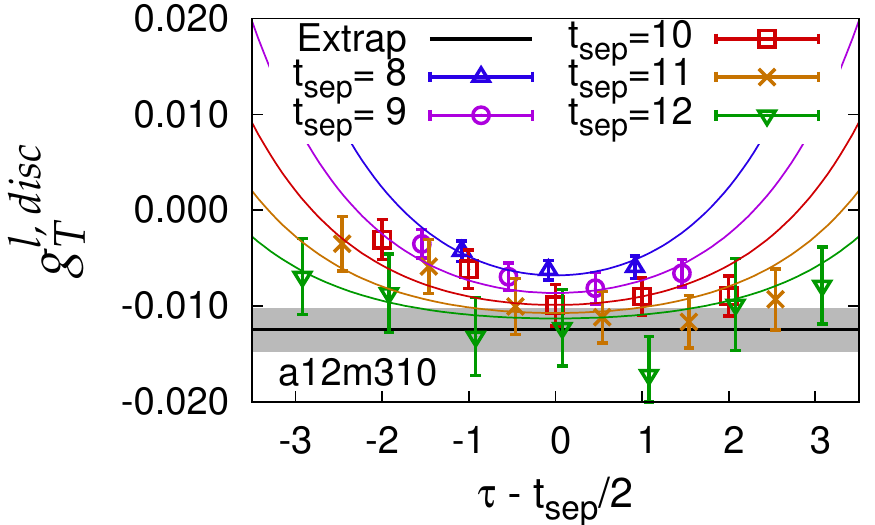}
    \includegraphics[width=0.33\linewidth]{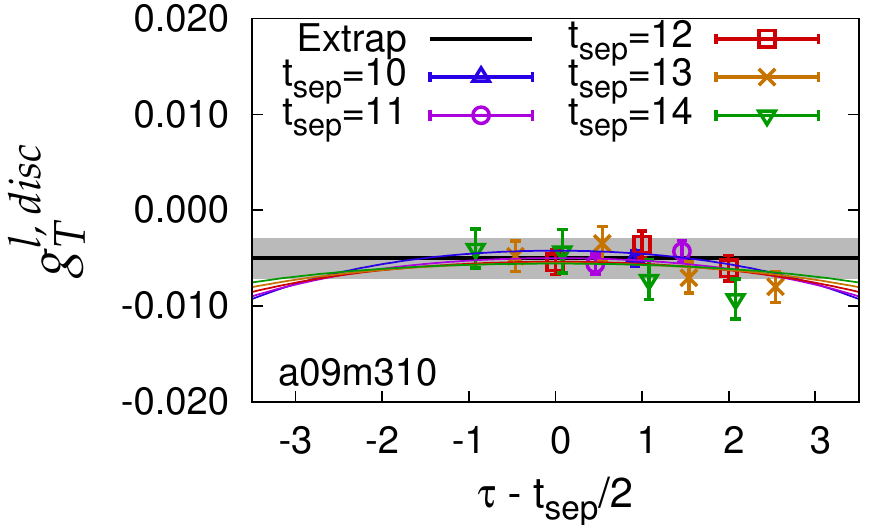}
    \includegraphics[width=0.33\linewidth]{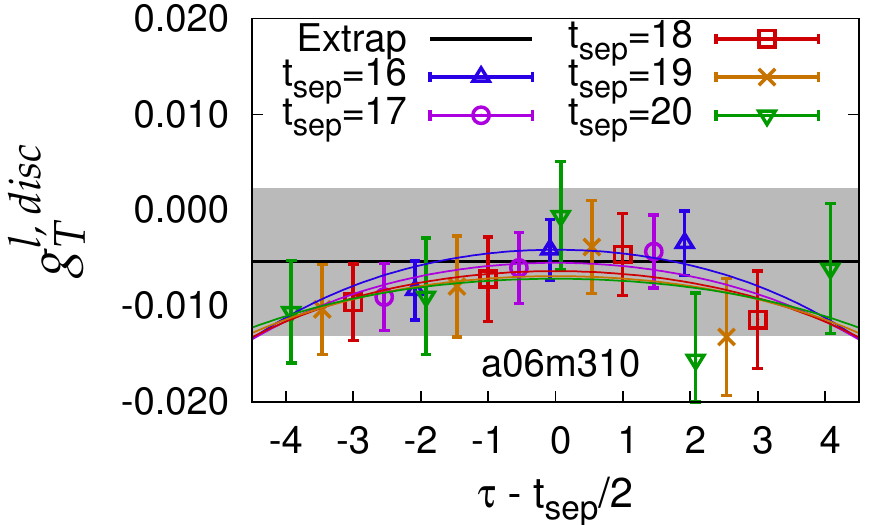}
    }
  \vspace{0.04\linewidth}
   \subfigure{
    \includegraphics[width=0.33\linewidth]{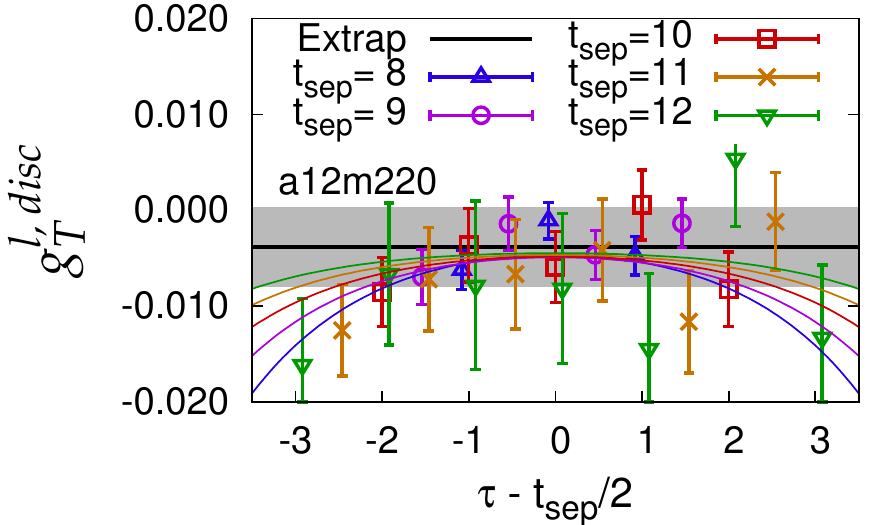}
    }
  \end{flushleft}
  \vspace{-0.07\linewidth}
\caption{Fits, using Eq.~\protect\eqref{eq:2pt_3pt}, to isolate the
  excited state contribution in the light quark disconnected diagram,
  $g_T^{\rm l,disc}$, are shown for the four ensembles analyzed.  The
  solid black line and the grey band are the ground state estimate and
  error. The data and results of the fit for different $t_\text{sep}$
  are also shown.
  \label{fig:ext_t_disc_l}}
\end{figure*}

\begin{figure*}[tb]
  \begin{flushleft}
   \subfigure{
    \includegraphics[width=0.33\linewidth]{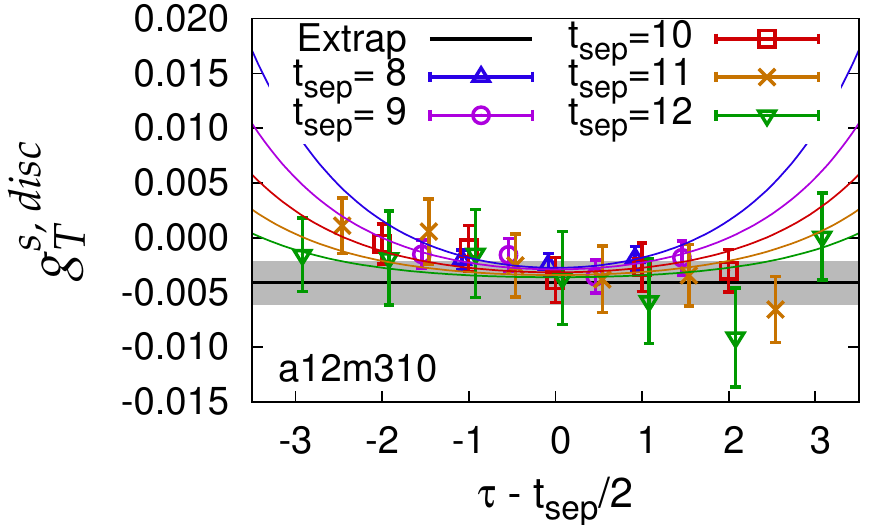}
    \includegraphics[width=0.33\linewidth]{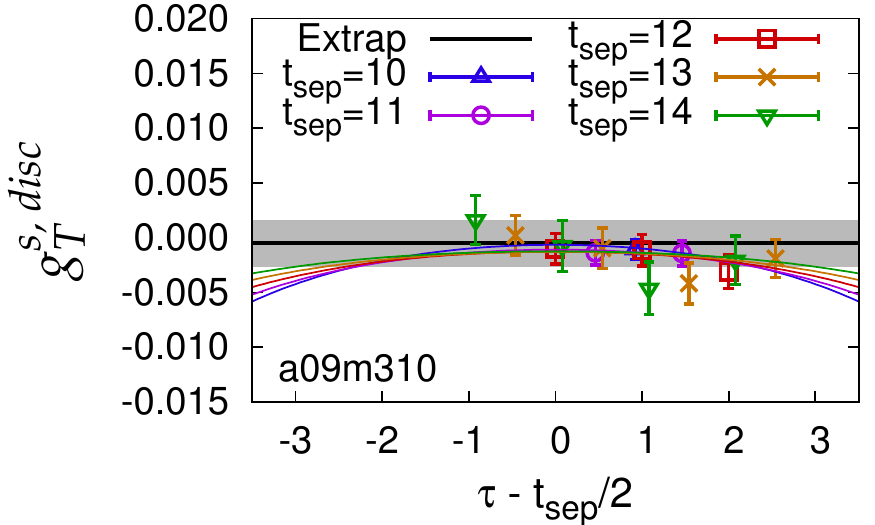}
    \includegraphics[width=0.33\linewidth]{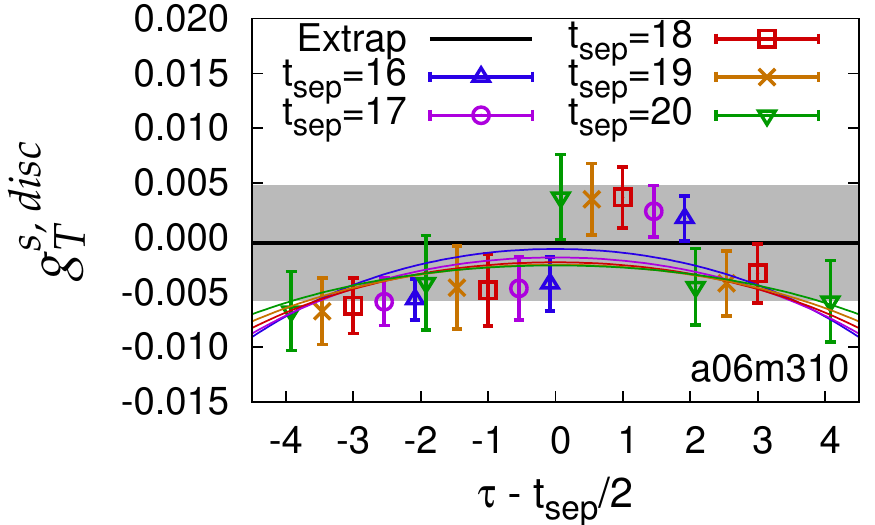}
    }
  \vspace{0.04\linewidth}
   \subfigure{
    \includegraphics[width=0.33\linewidth]{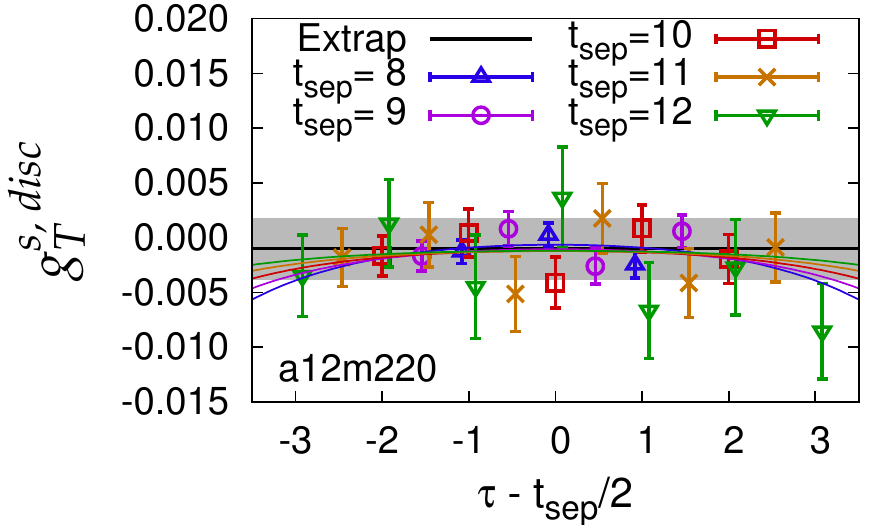}
    \includegraphics[width=0.33\linewidth]{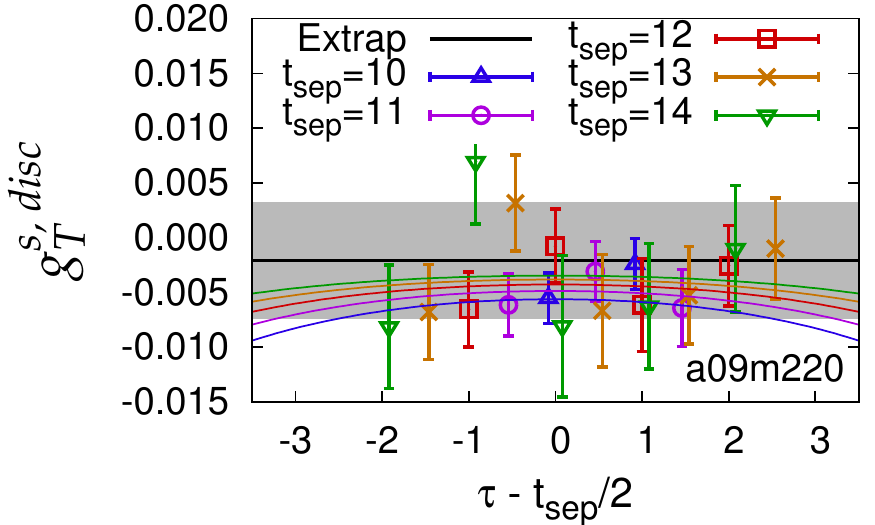}
    \hfill
    }
  \end{flushleft}
  \vspace{-0.07\linewidth}
\caption{Fits, using Eq.~\protect\eqref{eq:2pt_3pt}, to isolate the
  excited state contribution in the strange quark disconnected
  diagram, $g_T^{\rm s,disc}$, are shown for the five ensembles
  analyzed.  The solid black line and the grey band are the ground
  state estimate and error. The data and results of the fit for
  different $t_\text{sep}$ are also shown.
  \label{fig:ext_t_disc_s}}
\end{figure*}

We use the fit ansatz given in Eq.~\eqref{eq:2pt_3pt}, $i.e.$ the same
{\it two-state-fit} method used to fit the data for the connected
three-point diagrams, to extract the ground state results for the
disconnected contribution.
The data and the results of the fit for the light and strange quark
loop on the {\it a12m310} ensemble are shown in
Figs.~\ref{fig:ext_t_disc_l} and~\ref{fig:ext_t_disc_s},
respectively. We find significant contribution from excited states
only on the {\it a12m310} ensemble for light quark disconnected
diagram --- it is large for $t_{\rm sep}=8$, but by $t_{\rm sep}=12$
the data agree with the final extrapolated value.  The peculiar
pattern seen in the {\it a06m310} ensemble is most likely due to the
small number (100 [200]) of configurations analyzed as given in
Table~\ref{tab:params_disc}.

Including the disconnected diagrams also requires calculating their 
contribution to the renormalization constants in the 
RI-sMOM scheme.  We have not done this due to the poor signal in 
disconnected diagrams and use the same renormalization factor as
calculated for the connected diagrams.  In perturbation theory, the
disconnected diagrams come in at higher order, so their contributions
are expected to be small. Furthermore, the disconnected quark
loop contributions themselves are very small for the nucleon tensor
charges, so we expect the impact of the small difference in the
renormalization factor due to neglecting the disconnected piece in
$Z_T$ will change the final estimate by much less than the statistical
error quoted in Table~\ref{tab:res}.
The final renormalized results, with this caveat, are given in
Table~\ref{tab:res-renorm}.

There are two ways in which we can study the quark mass dependence of
the disconnected contribution. First, by comparing the strange with
light quark loop contributions we note that the estimates on all four
ensembles increase as the quark mass is decreased.  The second is to
compare the estimates on the {\it a12m310} and {\it a12m220}
ensembles.  Unfortunately, the statistical errors in the latter are
too large to draw a conclusion, even though we used the largest number
of random sources per configuration for this study. Our conclusion is
that a higher statistics study is needed to quantify the quark mass
dependence and reduce the overall error in the disconnected
contribution so that a reliable continuum extrapolation can be made.

The authors in Ref.~\cite{Abdel-Rehim:2013wlz} found that the
disconnected contribution to the nucleon tensor charge is consistent
with zero on a $N_f=2+1+1$ twisted mass fermion ensemble at
$a=0.082(4)\fm$ and $M_\pi = 370\MeV$. While a direct comparison with
our results would be meaningful only after both results have been
extrapolated to the continuum and physical pion mass limit, we note
that our estimates are also consistent with zero for all ensembles
with the strange quark loop, and in two of the four cases of light
quark loops.

\begin{figure*}[tb]
\includegraphics[width=0.96\linewidth]{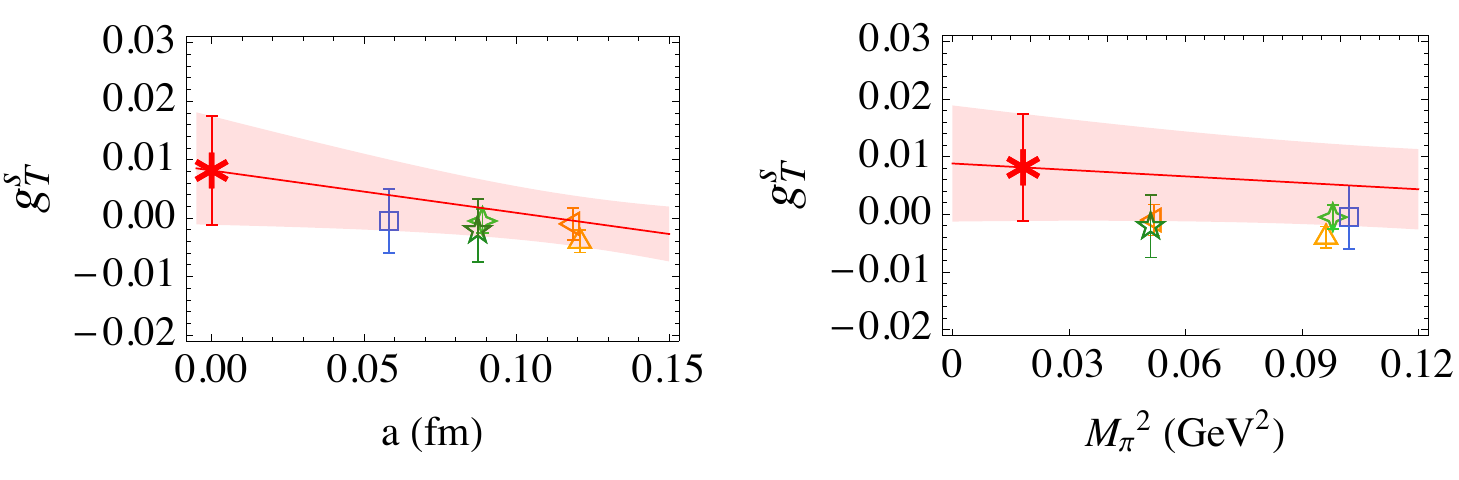}
\caption{Fits using Eq.~\protect\eqref{eq:extrap} to obtain the result
  in the continuum limit ($a\rightarrow 0$) and at the physical pion
  mass ($M_\pi \rightarrow M_{\pi^0}^{\text{phys}}$) of the strange
  quark disconnected contribution. A finite volume study was not
  carried out for the disconnected contribution. Rest is the same as
  in Fig.~\protect\ref{fig:chiral_gT_compare}.
  \label{fig:disc_s_extrap}}
\end{figure*}

Given that the estimates of the disconnected contribution with light
quark loops are small compared to connected part, have large errors,
and have been obtained on only four ensembles, we do not include them
in estimates of the isoscalar charges $g_T^{(u,d)}$. Instead, we take
the largest value $0.0121$ on the {\it a12m310} ensemble and use it as
an estimate of the systematic error associated with neglecting the
disconnected piece.  This error is added in quadrature to the overall
error in the connected estimate.  The disconnected contribution with
strange quark loops is even smaller but we keep it since it does not
have a connected piece and we can perform a reasonable extrapolation
in the lattice spacing and the quark mass as shown in
Fig.~\ref{fig:disc_s_extrap}, and get
\begin{equation}
g_T^{\rm s,disc} = 0.008(9) \,,
\label{eq:gTs}
\end{equation}
with a $\chi^2/\text{dof} = 0.29$ for $\text{dof}=2$. Bounding
$g_T^{s}$ is important for the analysis of the neutron EDM, especially
if the chirality flip is controlled by the Higgs Yukawa coupling. In
those BSM scenarios, the contribution of $g_T^{s}$ would be enhanced
by the ratio of quark masses $m_s/m_{u,d}$ ($i.e.$, proportional to
the coupling of a ``Higgs'' to quarks), relative to $g_T^{u}$ and
$g_T^{d}$.  Using these estimates, the analysis of the contribution of
the quark EDMs to the neutron EDM in presented in
Section~\ref{sec:final_nEDM}.

\begin{table*}
\begin{ruledtabular}
\begin{tabular}{c|cc|cccc|c}
ID & $t_\text{sep}/a$ & $\tau/a$ & $N_{\text{conf}}^{disc,l}$  & $N_{\text{conf}}^{disc,s}$ & $N_{\text{LP}}^{disc,l}$ & $N_{\text{LP}}^{disc,s}$ & $N_\text{LP}/N_\text{HP}$  \\
\hline
a12m310 & $\{ 8\sim 14\}$ & $3,4, \cdots, 11$        & 1013 & 1013 &  5000 & 1500 & 30  \\
a12m220 & $\{ 8\sim 14\}$ & $3,4, \cdots, 11$        & 958  & 958  & 11000 & 4000 & 30  \\
a09m310 & $\{10\sim 16\}$ & $6, 7, 8, 9$             & 1081 & 1081 &  4000 & 2000 & 30  \\
a09m220 & $\{10\sim 16\}$ & $5, 6, 7, 8, 9$          & ---  & 200  & 10000 & 8000 & 50  \\
a06m310 & $\{16\sim 24\}$ & $6,8,10,\cdots,18$       & 100  & 200  & 10000 & 5000 & 50  \\
\end{tabular}
\end{ruledtabular}
\caption{The parameters used in the study of the disconnected diagrams. 
  The source-sink time separations analyzed ($t_\text{sep}$), the
  time slices ($\tau$) on which the operator is inserted as explained in the text, the number of
  configurations analyzed ($N_\text{conf}$) and the number of random noise
  sources ($N_\text{LP}$) used on each configuration.
  Here $\{A\sim B\}$ denotes the set of consecutive integers from $A$ to $B$. 
%
%
 The number $N_\text{LP}/N_\text{HP}$ gives
 the ratio of the number of low to high precision calculations
 done. The LP criteria for stopping the Dirac matric inversion was set
 to $r_{\rm LP} = 0.005$.  For the associated two-point function
 calculation, we used AMA with 64 LP and 4 HP measurements on each
 configuration and the results for the masses and amplitudes are given 
 in Table~\ref{tab:res2pt}. 
}
\label{tab:params_disc}
\end{table*}
%

\section{Nucleon Tensor Charges and Quark Electric Dipole Moment}
\label{sec:final_results}

In the previous Sections~\ref{sec:connected} and~\ref{sec:resultsD},
we discussed the calculation of the connected and disconnected
diagrams to the nucleon tensor charges. In this section we present our
final results for the nucleon tensor charges and the constraints they
put on the quark EDM couplings using the current bound on the neutron
EDM.
%

\subsection{Nucleon Tensor Charge}
\label{sec:final_gT}

The isovector tensor charge $g_T^{u-d}$, needed to probe novel tensor
interactions at the TeV scale in the helicity-flip part of neutron
decays, does not get any contributions from the disconnected diagram
in the iso-spin symmetric limit that we are working under. We consider
the extraction of $g_T^{u-d}$ reliable because all systematics are
under control. In particular, we find (i) that the fit ansatz in
Eq.~\eqref{eq:2pt_3pt} converges, indicating that the excited state
contamination has been isolated. (ii) the data for the renormalization
constant in the RI-sMOM scheme shows a window in $q^2$ for which the
final estimates in the $\overline{\text{MS}}$ scheme at $2\GeV$ are
constant within errors as discussed in
Section~\ref{sec:renorm}. Finally (iii), the estimates on the nine
ensembles show little dependence on the lattice spacing, pion mass and
lattice volume as shown in Fig.~\ref{fig:con_extrap} and discussed in
Section~\ref{sec:connected}.

Our final estimate given in Eq.~\eqref{eq:conn_extrap2}, $ g_T^{u-d} =
1.020(76)$, is in good agreement with other lattice calculations by
the LHPC ($N_f=2+1$ HEX smeared clover action, domain wall action, and
domain wall-on-asqtad actions)~\cite{Green:2012ej}, RBC/UKQCD
($N_f=2+1$ domain wall fermions~\cite{Aoki:2010xg}, ETMC ($N_f=2+1+1$
twisted mass fermions)~\cite{Alexandrou:2013wka,Alexandrou:2014wca,Abdel-Rehim:2015owa} and the RQCD ($N_f=2$
$O(a)$ improved clover fermions)~\cite{Bali:2014nma} as shown in
Fig.~\ref{fig:GlobalData}. A more detailed discussion of the
systematics in these calculations and control over them using the FLAG
quality criteria~\cite{FLAGqc} is given in the
Appendix~\ref{app:FLAG}.

An analysis of the extrapolation to the physical quark mass has also
been carried out by the LHPC~\cite{Green:2012ej} and
RQCD~\cite{Bali:2014nma} collaborations. They did not find significant
dependence on the lattice spacing and volume, so they extrapolate only
in the quark mass using linear/quadratic (LHPC) and linear (RQCD) fits
in $M_\pi^2$.  Their final estimates, $g_T^{u-d} = 1.038(11)(12)$
(LHPC) and $g_T^{u-d} = 1.005(17)(29)$ (RQCD) are consistent with
ours, but the size of our error is much larger.  This is due to a
combination of three factors in our calculation: (i) our determination
of renormalization constants have larger uncertainty; (ii) errors in
individual points are larger because they are the estimates in the
$t_{\rm sep} \to \infty$ limit obtained by extrapolating the data with
multiple $t_{\rm sep}$ using a two state ansatz; and (iii) we
extrapolate in all three variables using Eq.~\eqref{eq:extrap},
whereas LHPC and RQCD extrapolate only in $M_\pi^2$.  A fit to our
data versus only $M_\pi^2$, also shown in Fig~\ref{fig:conUD_extrap},
gives a similarly accurate estimate $g_T^{u-d} = 1.059(29)$ with a
$\chi^2/{\rm dof} = 0.3$.

\begin{figure}[tb]
    \includegraphics[width=0.8\linewidth]{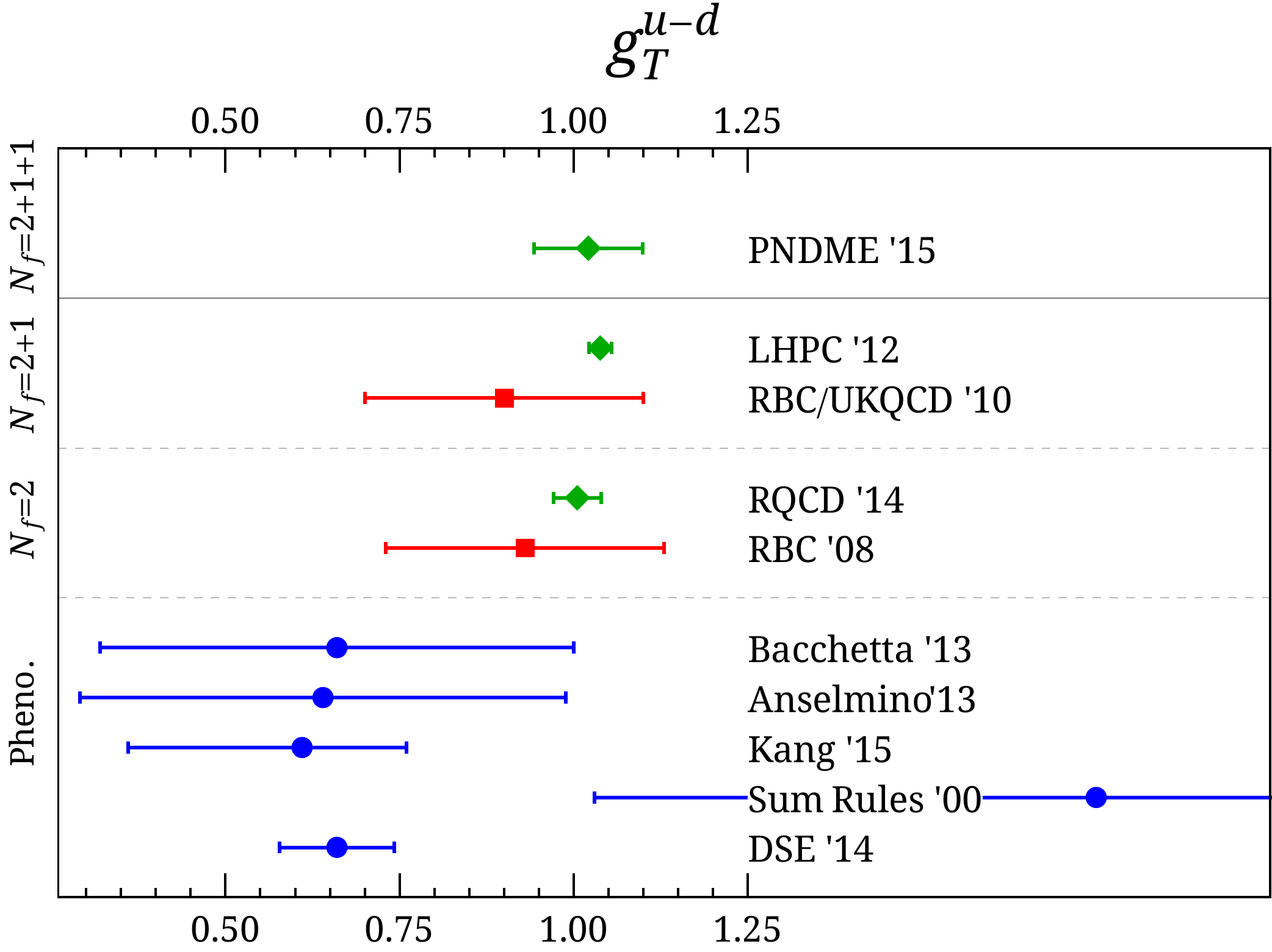}
\caption{A comparison between recent lattice QCD results for
  $g_T^{u-d}$ and estimates derived from model calculations and
  experimental data.  The published lattice QCD results are from
  LHPC'12~\cite{Green:2012ej}, RBC/UKQCD'10~\cite{Aoki:2010xg},
  RQCD'14~\cite{Bali:2014nma} and RBC'08~\cite{Lin:2008uz}. Lattice
  estimates with reasonable control over excited state contamination and
  extrapolation to the physical pion mass and the continuum limit are
  shown in green.  Estimates from models and phenomenology are from
  Bacchetta'13~\cite{Bacchetta:2012ty},
  Anselmino'13~\cite{Anselmino:2013vqa}, Kang'15~\cite{Kang:2015msa},
  Sum Rules'00~\cite{Pospelov:1999rg},
  DSE'14~\cite{Pitschmann:2014jxa}.
\label{fig:FLAG}}
\end{figure}

A comparison between recent lattice QCD results for $g_T^{u-d}$ and
estimates derived from model calculations and experimental data are
summarized in Fig.~\ref{fig:FLAG}.\footnote{A similar comparison is
presented in Ref.~\cite{Courtoy:2015haa}.} The lattice estimates
show consistency and little sensitivity to the number of flavors,
$i.e.$, $N_f=2$ or $2+1$ or $2+1+1$, included in the generation of
gauge configurations. The errors in model and phenomenological
estimates (integral over the longitudinal momentum fraction of the
experimentally measured quark transversity distributions) are
large. Only the Dyson-Schwinger estimate (DSE'14) has comparable
errors and lies about 4$\sigma$ below the lattice QCD estimates.

To summarize, even with a very conservative error estimate, our result
$ g_T^{u-d} = 1.020(76)$, meets the target uncertainty of $\sim 10\%$
required to bound novel tensor interactions using measurements of the
helicity flip part of the neutron decay distribution in experiments
planning to reach $10^{-3}$ accuracy. Our goal for the future is to
reduce the error in $g_S$, which currently is $\sim 30\%$ for the
data sets presented in this work, to the same level.

\subsection{Quark Electric Dipole Moment}
\label{sec:final_nEDM}

The quark EDM contributions to the neutron EDM, $d_n$, are given by 
\begin{equation}
    d_n  = d_u ~g_T^{u} + d_d ~g_T^{d} + d_s ~g_T^{s}
\label{eq:nEDM}
\end{equation}
where the low-energy effective couplings $d_u$, $d_d$ and $d_s$
encapsulate the new CP violating interactions at the TeV scale.  The
goal of the analysis, knowing the charges $g_T^q$ and a bound on $d_n$,
is to constrain the couplings $d_q$ and, in turn, BSM theories. 

The calculation of the connected contribution to the $g_T^q$ has been
discussed in Section~\ref{sec:final_gT}. Estimates of the disconnected
contribution were discussed in Section~\ref{sec:resultsD}. Including
the largest value ($0.0121$ obtained on the {\it a12m310} ensemble) as
a systematic error, our final results in the $\overline{\text{MS}}$
scheme at $2\GeV$ for the nucleon charges that get contributions from
the disconnected diagrams are:
\begin{align}
  g_T^{u} &= 0.774(66)\,, \nonumber \\
  g_T^{d} &= -0.233(28)\,, \nonumber \\
  g_T^{u+d} &= 0.541(67)\,.
\label{eq:gTfinal}
\end{align}
%
%
Note that incorporating the disconnected contribution as a systematic
error increases the errors marginally as can be seen by comparing
estimates in Eq.~\eqref{eq:gTfinal} with those in
Eq.~\eqref{eq:conn_extrap}. Results for the neutron tensor
charges are obtained by using the iso-spin symmetry, $i.e.$, by
interchanging the labels $u \leftrightarrow d$.

These final estimates are significantly smaller in magnitude than the
quark model values, $g_T^{u}=4/3$ and $g_T^{d}=-1/3$, but consistent
with estimates derived from model calculations and experimental data
summarized in Table~\ref{tab:othergT}.\footnote{The effect of the
  choice of scale $\mu$ is illustrated by converting the results in
  Ref.~\cite{Bacchetta:2012ty} from $\mu=1\GeV$ to $\mu=2\GeV$ using
  two-loop running with the anomalous dimensions taken from
  Refs.~\cite{Sturm:2009kb,Gracey:2000am}.} 
The three phenomenological estimates
Bacchetta'13~\cite{Bacchetta:2012ty},
Anselmino'13~\cite{Anselmino:2013vqa}, and Kang'15~\cite{Kang:2015msa}
give consistent but lower estimates for $g_T^{u}$ and $g_T^{d}$ with
$g_T^{u-d} \sim 0.65$.  Similarly, taking the errors at face value,
the Schwinger-Dyson estimate is $\sim 4\sigma$ below the lattice QCD
results. A recent reevaluation of the calculation of tensor charges
using QCD sum rules with input from lattice QCD has been reported
in~\cite{Fuyuto:2013gla,Hisano:2012sc}.  Their estimates in the
$\overline{\text{MS}}$ scheme at $1\GeV$ are $g_T^d = 0.79$ and $g_T^u
= -0.20$, each with $\approx 50\%$ uncertainty. Run to $2$~GeV, these
estimates would decrease by $\approx 10\%$ in magnitude. These results are
consistent with ours given in~\eqref{eq:gTfinal} but place less
stringent constraints on the neutron EDM and BSM theories due to the
larger uncertainty.

\begin{table*}
\renewcommand{\arraystretch}{1.4} 
\begin{tabular}{c|c|c|c|c}
\hline\hline
                                          & $g_T^{d}$   & $g_T^{u}$     & $g_T^{s}$     & $\mu$         \\ \hline
This study                                & $-0.23(3)  $& $0.77(7) $    & $0.008(9) $   & 2 GeV         \\ \hline
Quark model                               & $-1/3      $& $4/3     $    & --            & --             \\
QCD Sum Rules~\cite{Pospelov:1999rg}      & $-0.35(17) $& $1.4(7)  $    & --            & ?             \\
Dyson-Schwinger~\cite{Pitschmann:2014jxa} & $-0.11(2)  $& $0.55(8) $    & --            & 2 GeV         \\
Transversity 1~\cite{Bacchetta:2012ty}    & $-0.18(33) $& $0.57(21)$    & --            & $\sim$ 1 GeV  \\
Transversity 1~\cite{Bacchetta:2012ty}    & $-0.16(30) $& $0.51(19)$    & --            & 2 GeV  \\
Transversity 2~\cite{Anselmino:2013vqa}   & $-0.25(20) $& $0.39(15)$    & --            & $\sim$ 1 GeV  \\
Transversity 3~\cite{Kang:2015msa}        & $-0.22{}^{+0.14}_{-0.08}$ & $0.39{}^{+0.07}_{-0.11}$  & --     & $3.2$ GeV  \\
\hline\hline
\end{tabular}
\caption{A comparison of our lattice estimates of $g_T^{d}$ and $g_T^{u}$ of the proton with those from
  different models and phenomenology. The ``Transversity 1'' estimate is given both at
  the original scale at which it was evaluated ($\sim 1$ GeV) and
  after running to $2$~GeV to show the magnitude of the scaling
  effect. The symbol ``?'' in the last column indicates that the scale
  at which the calculation is done is undetermined.  }
\label{tab:othergT}
\end{table*}

\begin{figure}[tb]
    \includegraphics[width=0.96\linewidth]{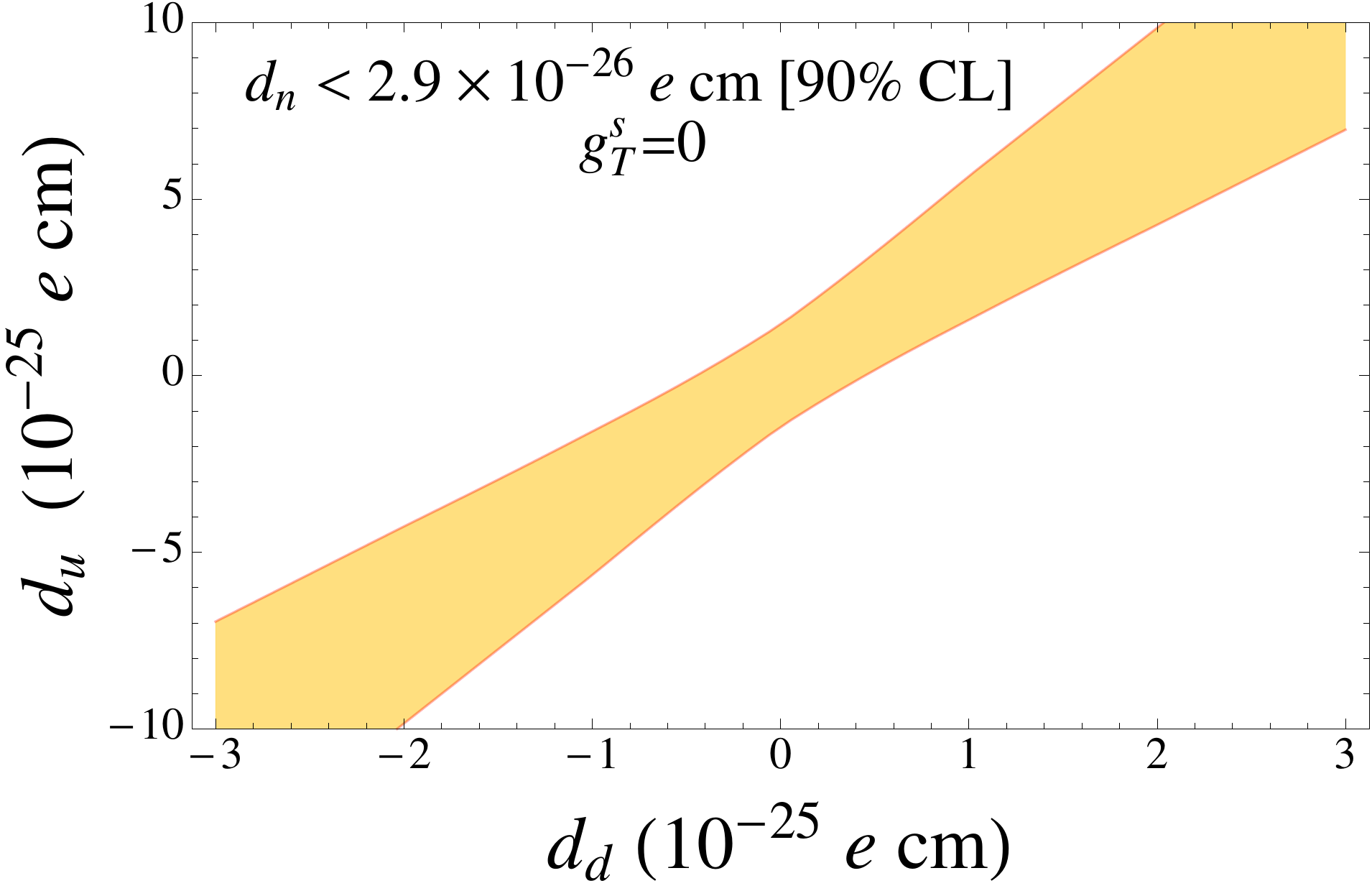}
\caption{Bounds on the couplings $d_{u,d}$ for the case
  $g_T^s = 0$.  Estimates used for
  $g_T^u$ and $g_T^d$ are given in
  Eq.~\protect\eqref{eq:gTfinal}.
  \label{fig:bounds}}
\end{figure}

Assuming that only the EDMs of the $u$, $d$, and $s$ quarks
contribute to the neutron EDM via Eq.~\eqref{eq:nEDM} and the values
of $g_T^{u,d,s}$ are given by Eqs.~\eqref{eq:gTfinal} and
\eqref{eq:gTs}, one can put bounds on the $d_{u,d,s}$.  Using the
current estimate $|d_N| < 2.9 \times 10^{-26}$~$e$~cm (90\%
CL)~\cite{Baker:2006ts}, 1-sigma slab priors for $g_T^{u}$ and
$g_T^{d}$ given in Eq.~\eqref{eq:gTfinal}, and assuming $g_T^s = 0$,
we obtain the 90\% confidence interval bounds for $d_{u}$ and $d_{d}$
shown in Fig.~\ref{fig:bounds}. Note that $d_s$ is not constrained
since $g_T^s$ is consistent with zero.

Using these estimates of $g_T^{u,d,s}$, 
we have analyzed the consequences on split SUSY models, in which the
quark EDM is the leading contribution in~\cite{Bhattacharya:2015esa}.
Our goal for the future is to improve the estimates presented here and
develop the lattice methodology to include the contributions of the
quark chromo electric dipole moment operator.

%

\section{Conclusions}
\label{sec:conclusions}

We have presented a high statistics study of the isovector and
isoscalar tensor charges of the nucleon using clover-on-HISQ lattice
QCD. We calculate both the connected and disconnected diagrams
contributing to these charges. The analysis of nine ensembles covering
the range $0.12-0.06\fm$ in lattice spacing, $M_\pi = 130-320$~MeV in
pion mass, and $M_\pi L = 3.2 - 5.4$ in lattice volume allowed us to
control the various sources of systematic errors.  We show that
keeping one excited state in the analysis allows us to isolate and
mitigate excited state contamination. The renormalized estimates of the
various tensor charges show small dependence on the lattice volume,
lattice spacing and the light quark mass. These results can, therefore, 
be extrapolated reliably to the physical point.

Our final estimate for the tensor charge $g_T^{u-d} = 1.020(76)$ is in
good agreement with previously reported estimates. The signal in the
calculation of the disconnected diagrams is weak in spite of using
state-of-the-art error reduction techniques. The value is small and we
bound its contribution to light quark charges $g_T^{u}$ and $g_T^{d}$.
The signal for strange disconnected loop is even smaller, however in
this case we are able to extrapolate the results to the continuum
limit and find $g_T^s = 0.008(9)$.  Using these estimates and the
current bound on the neutron electric dipole moment, we carry out a
first lattice QCD analysis of the constraints on the strengths of the
up, down and strange quark electric dipole moments. The impact of
these constraints on the viability of split SUSY models, in which the
quark EDM is the leading contribution to the neutron EDM, is carried out
in~\cite{Bhattacharya:2015esa}.

\begin{acknowledgments}
We thank the MILC Collaboration for providing the 2+1+1 flavor HISQ
lattices used in our calculations.  Simulations were carried out on
computer facilities of (i) the USQCD Collaboration, which are funded
by the Office of Science of the U.S. Department of Energy, (ii) the
Extreme Science and Engineering Discovery Environment (XSEDE), which
is supported by National Science Foundation Grant No. ACI-1053575,
(iii) the National Energy Research Scientific Computing Center, a DOE
Office of Science User Facility supported by the Office of Science of
the U.S. Department of Energy under Contract No. DE-AC02-05CH11231; and 
(iv) Institutional Computing at Los Alamos National Laboratory. 
The calculations used the Chroma software
suite~\cite{Edwards:2004sx}. This material is based upon work supported by 
the U.S. Department of Energy, Office of Science of High Energy Physics under 
Contract No.~DE-KA-1401020 and the LANL LDRD program. The work of H.W.L and
S.D.C was supported by DOE Grant No.~DE-FG02-97ER4014. We thank Gunnar Bali, 
Martha Constantinou and Jeremy Green for providing their latest data, and 
Emanuele Mereghatti for discussions on the chiral extrapolation. We thank 
Constantia Alexandrou, Gunnar Bali, Tom Blum, Shigemi Ohta, Dirk Pleiter 
and the LHP collaboration for discussions on the FLAG analysis. 
\end{acknowledgments}

\vspace{4 in}

\appendix
\section{Systematics in the calculation of the isovector nucleon tensor charge}
\label{app:FLAG}

In Table~\ref{tab:FLAGgT}, we give a summary, in the FLAG
format~\cite{FLAG}, of the level of control over various systematics
in the calculation of the isovector tensor charge of the nucleon
using simulations of lattice QCD with $N_f=2$, $2+1$ and $2+1+1$
flavors. Note that a community wide consensus on applying the FLAG
criteria to matrix elements within nucleon states does not yet
exist. By performing this analysis, we wish to emphasize that the
agreement between various calculations of $g_T^{u-d}$ has reached a
level of precision that calls for a FLAG like analysis.

\begin{figure}[h]
    \includegraphics[width=0.96\linewidth]{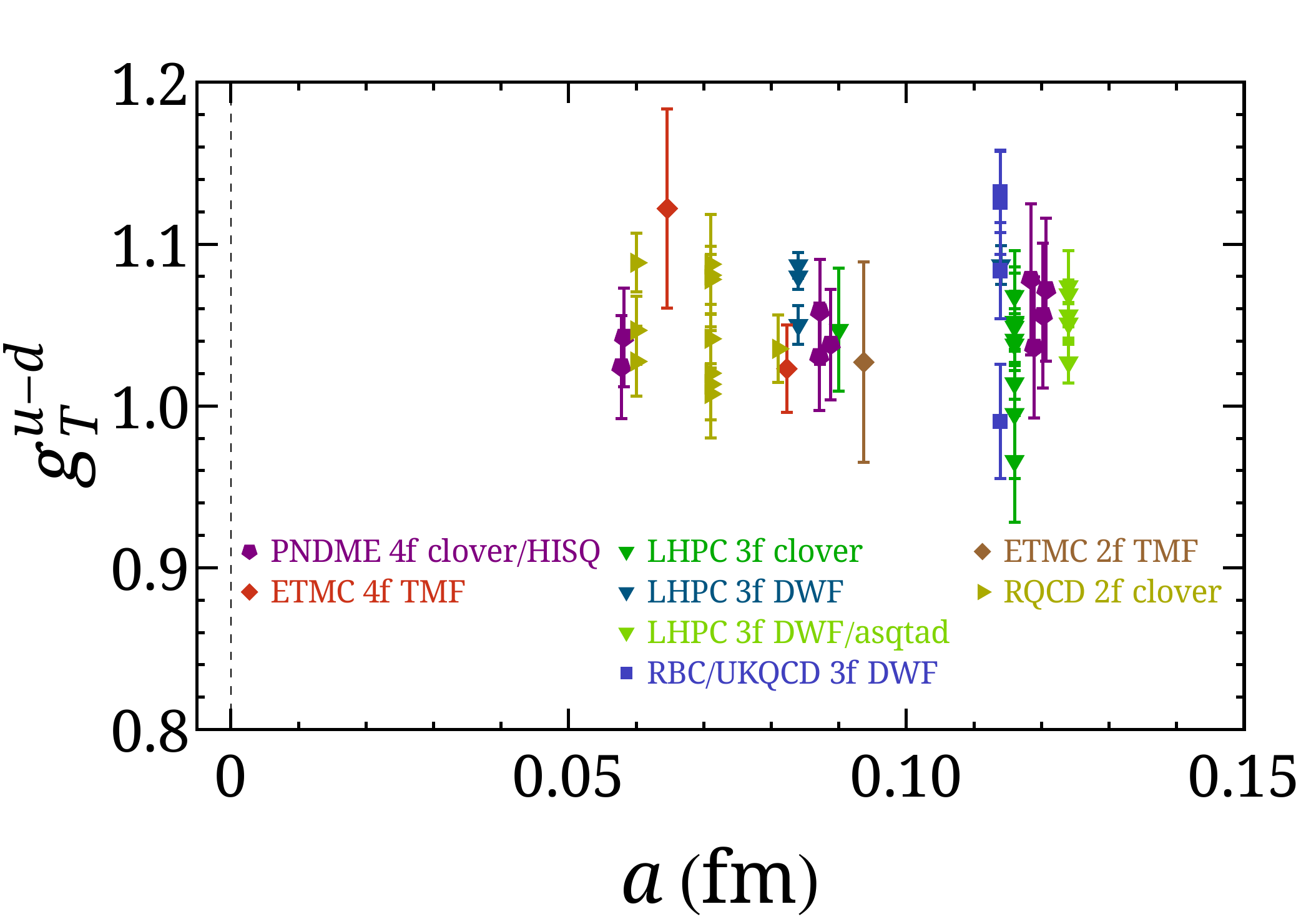}
    \includegraphics[width=0.96\linewidth]{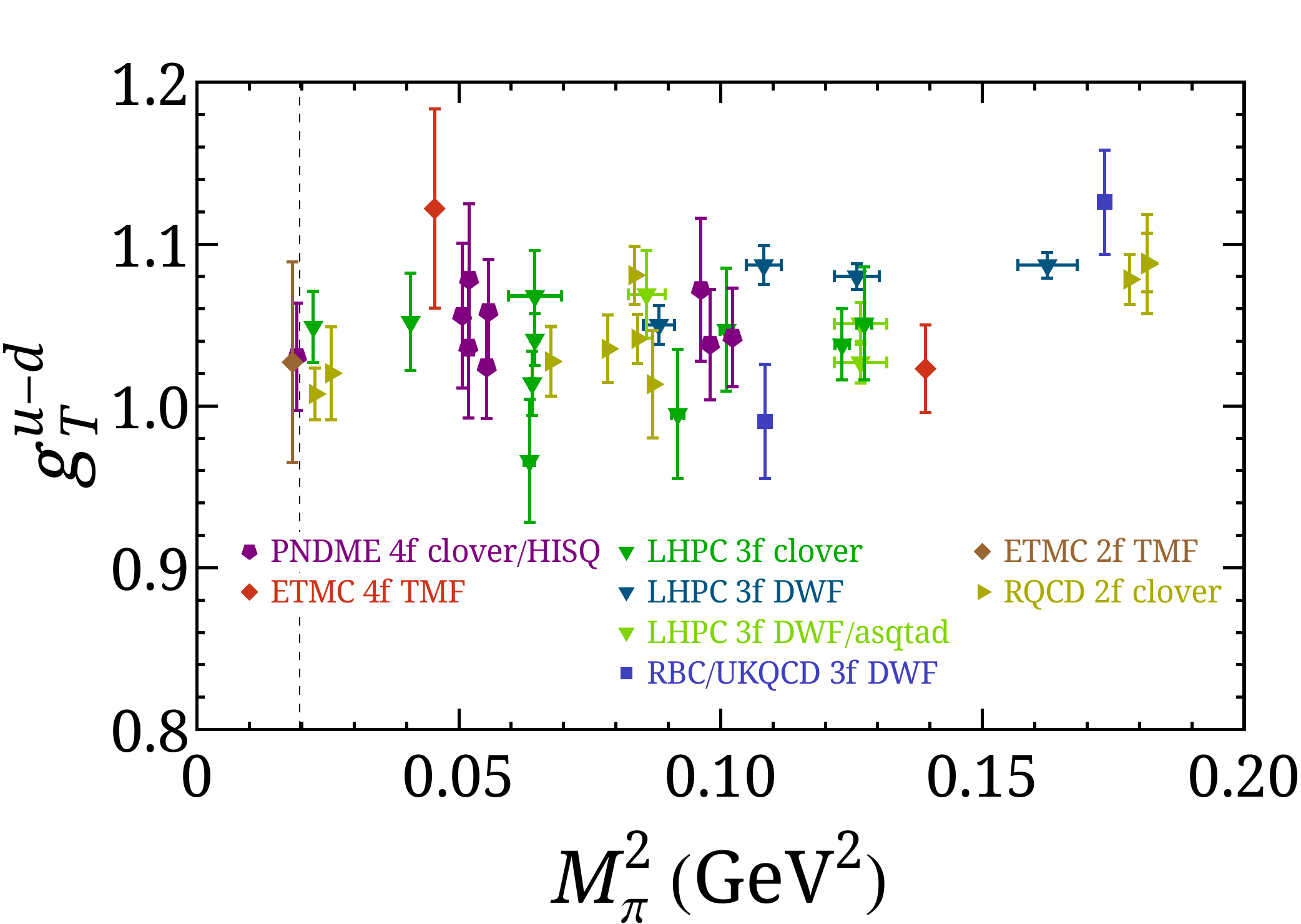}
    \includegraphics[width=0.96\linewidth]{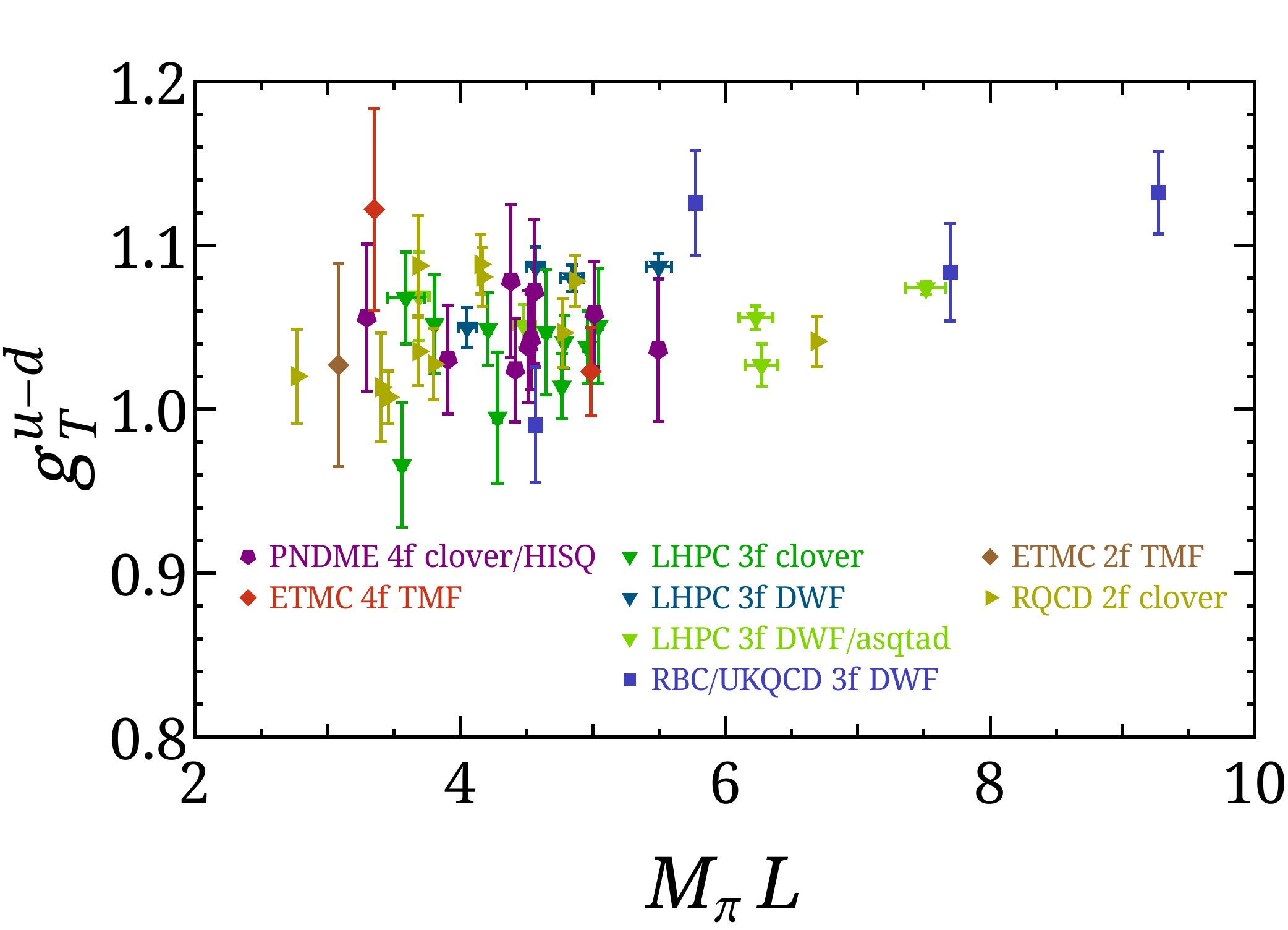}
\caption{Estimates of $g_T^{u-d}$ from lattice QCD for $N_f=2$, $2+1$ and $2+1+1$ 
flavors from the PNDME'15 (this work), ETMC'15~\cite{Alexandrou:2013wka,Alexandrou:2014wca,Abdel-Rehim:2015owa} 
  LHPC'12~\cite{Green:2012ej}, RBC/UKQCD'10~\cite{Aoki:2010xg},
  and RQCD'14~\cite{Bali:2014nma} 
collaborations.  These data show little sensitivity to $a$ (top),
$M_\pi^2$ (middle), $M_\pi L$ (bottom) and on whether the strange and
charm quarks are included in the generation of the lattice ensembles
or on the lattice action used. The vertical dashed line in the middle
panel marks the physical pion mass $M_\pi = 135$~MeV.
\label{fig:GlobalData}}
\end{figure}

The systematics covered by the FLAG criteria are also encountered in
the calculation of matrix elements within baryon states. We,
therefore, follow the same quality criteria for the publication
status, chiral extrapolation, finite volume effects, and
renormalization as defined by FLAG~\cite{FLAGqc} and define an
additional criterion, excited state contamination, that is relevant to
the calculations of matrix elements within nucleon states. For the
criterion ``continuum extrapolation'' we relax the requirement of an
extrapolation provided the data meet the rest of the requirements: do
not warrant an extrapolation, and a reasonable estimate of the
uncertainty is provided. We also do not require that the action and
the operators are $O(a)$ improved.
\begin{itemize}
\item Publication status: \\  
A \enspace published or plain update of published results \\
P \enspace preprint \\
C \enspace conference contribution
\item Chiral extrapolation:\\ 
\good \enspace $M_{\pi,\text{min}}< 200$~MeV  \\
\soso \enspace $200\text{ MeV}\le M_{\pi,\text{min}} \le 400$~MeV \\
\bad \enspace $400\text{ MeV} < M_{\pi,\text{min}}$ \\
\vspace{30pt}
\item Continuum extrapolation:\\ 
\good \enspace 3 or more lattice spacings, at least 2 points below 0.1~fm \\ 
\soso \enspace 2 or more lattice spacings, at least 1 point below 0.1~fm \\ 
\bad \enspace otherwise \\
\item Finite-volume effects:\\ 
\good \enspace $M_{\pi,\text{min}} L > 4$ or at least 3 volumes \\
\soso \enspace $M_{\pi,\text{min}} L > 3$ and at least 2 volumes \\
\bad \enspace otherwise\\
\item Renormalization:\\  
\good \enspace nonperturbative\\
\soso \enspace 1-loop perturbation theory or higher with a reasonable estimate of truncation errors\\
\bad \enspace otherwise \\
%
\item Excited State:\\ 
\good \enspace $t_\text{sep, max} > 1.5$~fm or
  at least 3 source-sink separations, $t_{\rm sep}$, investigated 
  at each lattice spacing and at each $M_{\pi}$. \\ 
\soso \enspace At least 2 source-sink separations with $ 1.2\ {\rm fm}\ \le
  t_\text{sep, max} \le 1.5\ {\rm fm}$ at at least one $M_{\pi}$ at each lattice spacing. \\ 
\bad \enspace otherwise\\
\end{itemize}
Plots of the data, summarized in Table~\ref{tab:FLAGgT}, as a function of
$a$, $M_\pi^2$ and $M_\pi L$ are shown in Fig.~\ref{fig:GlobalData}. 
One observes very little sensitivity to these three variables and
on the number of fermion flavors or the lattice action used.

\begin{table*}[h]
\begin{tabular*}{\textwidth}{l@{\extracolsep{\fill}}c lllllll l}
 Collaboration & Ref. & 
 \slantlabel{publication status} &
 $N_f$ &
 \slantlabel{chiral extrapolation} &
 \slantlabel{continuum extrapolation}  &
 \slantlabel{finite volume}  &  
 \slantlabel{excited state}  &  
 \slantlabel{renormalization} &  
 $g_T$ \\[3pt] 
\hline \hline 
PNDME'15 & This work & P & 2+1+1 & \good
 & \good & \good & \good & \good & 1.020(76)\footnote{This estimate is
   obtained from a simultaneous fit versus $a$, $M_\pi^2$, and
   $e^{-M_\pi L}$ defined in Eq.~\eqref{eq:extrap} using data on nine
   clover-on-HISQ ensembles given in Table~\ref{tab:res-renorm}.} \\
 
ETMC'15 & \cite{Abdel-Rehim:2015owa} & C & 2+1+1 &
\bad & \bad & \good & \good & \good & 1.053(21)\footnote{The quoted
  estimate~\cite{Abdel-Rehim:2015owa} is from a single $M_\pi= 373$ MeV, $a =
  0.082$~fm and $N_f = 2+1+1$ maximally twisted mass ensemble. Three values of 
  $t_{\rm sep} \approx 1$, $1.15$, and $1.3$~fm are analyzed for handling
  excited state contamination. We quote their result from 
  the two-state fit. A second low
  statistics study on an ensemble with $M_\pi= 213$ MeV and
  $a = 0.064$ fm gave a consistent estimate.  }  \\

\hline LHPC'12 & \cite{Green:2012ej} & A & 2+1 & \good & \soso & \good
& \soso & \good & 1.038(11)(12)\footnote{The central value is from a two
  parameter chiral fit to just the coarse Wilson ensembles data. This agrees
  with a three parameter chiral fit to data from three different
  lattice actions simulated at different lattice spacings and with
  different volumes. Uncertainty due to
  extrapolation in the lattice spacing $a$ and the finite volume
  controlled by $M_{\pi}L$ is expected to be small.}  \\
 
RBC/UKQCD'10 & \cite{Aoki:2010xg} & A & 2+1 & \soso & \bad & \good &
\bad & \good & 0.9(2)\footnote{Result is based on simulations at
  one lattice spacing $1/a=1.73$ ~GeV using domain wall fermions. The
  statistics for the ensembles corresponding to the four pion masses
  simulated, $M_\pi = 329,\ 416,\ 550,\ 668$ MeV, were 3728, 1424, 392, 424
  measurements, respectively. A single $t_{\rm sep} = 1.39$~fm was used. 
  }  \\ \hline
%
%
RQCD'14 & \cite{Bali:2014nma} & A & 2 & \good & \good & \good & \soso
& \good & 1.005(17)(29)\footnote{The result of this clover-on-clover
  study is obtained using a fit linear in $M_\pi^2$ keeping data with
  $M_\pi^2 < 0.1$~GeV${}^2$ only.  Data do not show significant
  dependence on lattice spacing or lattice volume.  Excited state
  study is done on three of the eleven ensembles. Most of the data are
  with $t_{\rm sep} \sim 1$~fm.  The second error is an estimate of
  the discretization errors assuming they are $O(a^2)$ since $O(a)$
  improved operators with 1-loop estimates for the improvement
  coefficients are used in calculations done on $a=0.081$, $0.071$ and
  $0.06$ fm lattices.  Preliminary estimates presented by the QCDSF
  collaboration~\cite{Pleiter:2011gw} are superseded by this
  publication~\cite{QCDSF11}.  } \\
ETMC'15 & \cite{Abdel-Rehim:2015owa} & P & 2 & \good & \bad & \bad &
\good & \good & 1.027(62)\footnote{Result from a single ensemble of
  maximally twisted mass fermions with a clover term at $a=0.093(1)$
  fm, $M_\pi=131$~MeV and $M_\pi L \approx 3$.  To control excited
  state contamination, three values of $t_{\rm sep} \approx 0.94$,
  $1.1$ and $1.3$~fm are analyzed. We quote their value from the 
  $t_{\rm sep} \approx 1.3$~fm analysis.} \\
%
%
RBC'08 & \cite{Lin:2008uz} & A & 2 & \bad & \bad & \good & \bad &
\good & 0.93(6)\footnote{Results based on one lattice spacing $
  1/a=1.7$~GeV with the DBW2 domain wall action, three values of quark
  masses with $M_\pi = 493,\ 607,\ 695$ MeV, and $O(500)$
  measurements. Only one $t_{\rm sep}=10$ (1.14 fm) was simulated
  except at the lightest mass where $t_{\rm sep} =12 $ data was
  generated but used only as a consistency check as it has large
  errors. }\\
%

\hline
\end{tabular*}
\caption{
A summary of the control over various sources of systematic errors in lattice QCD calculations 
of the isovector tensor charge $g_T^{u-d}$ using the FLAG quality criteria~\protect\cite{FLAGqc} 
reproduced in this Appendix. Results from all collaborations quoted in this table 
have used nonperturbative methods for calculating the renormalization constants. 
\label{tab:FLAGgT} }
\end{table*}

\clearpage
%
\bibliography{ref} 

\end{document}